\documentclass[11pt,twoside,a4paper,cmspaper,final,collab]{cms-tdr}
\def\svnVersion{a30ae2a-D}\def\svnDate{2026/07/17}\def\cmsCernNoTag{CERN-EP-2026-201}\def\cmsCernDate{\today}\def\cmsMessage{Submitted to Physical Review D}
\begin{document}\cmsNoteHeader{HIG-25-003}

\newcommand{\ggZZ}{\ensuremath{\Pg\Pg \to \PZ\PZ}\xspace}
\newcommand{\jhugen}{\textsc{JHUGen}\xspace}
\newcommand{\mcfm}{\textsc{mcfm}\xspace}
\newcommand{\mela}{\textsc{mela}\xspace}
\newcommand{\WW}{\ensuremath{\PW\PW}\xspace}
\newcommand{\ZZ}{\ensuremath{\PZ\PZ}\xspace}
\newcommand{\VBF}{\ensuremath{\mathrm{VBF}}\xspace}
\newcommand{\VH}{\ensuremath{\PV\PH}\xspace}
\newcommand{\WH}{\ensuremath{\PW\PH}\xspace}
\newcommand{\ZH}{\ensuremath{\PZ\PH}\xspace}
\newcommand{\ggH}{\ensuremath{\Pg\Pg\PH}\xspace}
\newcommand{\ZX}{\ensuremath{\PZ+\PX}\xspace}
\newcommand{\Hboson}{\PH\xspace}
\newcommand{\Hell}{\ensuremath{\PH\to4\Pell}\xspace}
\newcommand{\GH}{\ensuremath{\Gamma_\PH}\xspace}
\newcommand{\Gref}{\ensuremath{\Gamma_0}\xspace}
\newcommand{\mH}{\ensuremath{m_{\PH}}\xspace}
\newcommand{\mell}{\ensuremath{m_{4\Pell}}\xspace}
\newcommand{\offshell}{off-shell\xspace}
\newcommand{\onshell}{on-shell\xspace}
\newcommand{\HVV}{\ensuremath{\PH\PV\PV}\xspace}
\newcommand{\kappaq}{\ensuremath{\kappa_{\PQq}}\xspace}
\newcommand{\barkappau}{\ensuremath{\overline{\kappa}_{\PQu}}\xspace}
\newcommand{\barkappad}{\ensuremath{\overline{\kappa}_{\PQd}}\xspace}
\newcommand{\barkappas}{\ensuremath{\overline{\kappa}_{\PQs}}\xspace}
\newcommand{\barkappac}{\ensuremath{\overline{\kappa}_{\PQc}}\xspace}
\newcommand{\barkappab}{\ensuremath{\overline{\kappa}_{\PQb}}\xspace}
\newcommand{\barkappaq}{\ensuremath{\overline{\kappa}_{\PQq}}\xspace}
\newlength\cmsFigWidth
\newlength\cmsFigWidthii
\ifthenelse{\boolean{cms@external}}{
    \providecommand{\cmsTable}[1]{#1}
    \providecommand{\CP}{\ensuremath{C\!P}\xspace}
    \providecommand{\cmsLeft}{upper\xspace}
    \providecommand{\cmsRight}{lower\xspace}
    \providecommand{\cmsLLeft}{Upper\xspace}
    \providecommand{\cmsRRight}{Lower\xspace}
    \setlength{\cmsFigWidth}{0.49\textwidth}
    \setlength{\cmsFigWidthii}{0.49\textwidth}
}{
    \providecommand{\cmsTable}[1]{\resizebox{\textwidth}{!}{#1}}
    \providecommand{\CP}{\ensuremath{CP}\xspace}
    \providecommand{\cmsLeft}{left\xspace}
    \providecommand{\cmsRight}{right\xspace}
    \providecommand{\cmsLLeft}{Left\xspace}
    \providecommand{\cmsRRight}{Right\xspace}
    \setlength{\cmsFigWidth}{0.65\textwidth}
    \setlength{\cmsFigWidthii}{0.8\textwidth}      
}

\cmsNoteHeader{HIG-25-003}
\title{Off-shell Higgs boson measurements: Yukawa couplings, self-coupling, compositeness, and width}

\date{\today}
\abstract{
Measurements of Higgs boson production in the off-shell region are presented, using the four-lepton decay channel. Data from proton-proton collisions at the CERN LHC, collected by the CMS experiment and corresponding to an integrated luminosity of 138\fbinv at a center-of-mass energy of 13\TeV, are utilized. The first direct test of composite Higgs boson models is performed, with a lower limit on the compositeness scale $\Lambda_\PH$ set at 870\GeV at the 95\% confidence level. Tests of gluon-fusion production within the standard model effective field theory framework are performed. The first constraint on the Higgs boson self-coupling in the off-shell region is obtained. By combining on- and off-shell measurements, the analysis sets the tightest constraints to date on light-quark Yukawa couplings, while relaxing assumptions such as the bound on the Higgs boson coupling to vector bosons, thereby providing more model-independent results. Constraints on the Higgs boson width are provided while accounting for a range of beyond-the-standard-model effects, including both light and heavy particles in the gluon-fusion production loop, as well as modified couplings to vector bosons. A combined analysis of the $\PH\to\PZ\PZ$ and $\PH\to\PW\PW$ channels is performed to improve sensitivity to the Higgs boson width, yielding $\Gamma_{\PH}=5.1^{+2.0}_{-1.8}\MeV$. The scenario of no off-shell Higgs boson production is excluded at a confidence level exceeding 5 standard deviations.
}

\hypersetup{
pdfauthor={CMS Collaboration},
pdftitle={Off-shell Higgs boson measurements: Yukawa couplings, self-coupling, compositeness, and width},
pdfsubject={CMS},
pdfkeywords={CMS, Yukawa couplings, substructure, Higgs boson, width, self-coupling}}

\maketitle

\section{Introduction}\label{sec:Introduction}
The standard model (SM) of particle physics postulates the existence of a field responsible for
the generation of the masses of the fundamental particles. The excitation of this field is known as the 
Higgs boson (\PH)~\cite{StandardModel67_1, Englert:1964et, Higgs:1964ia, Higgs:1964pj, Guralnik:1964eu,
StandardModel67_2, StandardModel67_3}.
It was observed in 2012~\cite{Aad:2012tfa,Chatrchyan:2012xdj,Chatrchyan:2013lba} 
with a mass (\mH) around 125\GeV~\cite{Aad:2015zhl,ATLAS_mass} and a width 
(\GH)~\cite{Khachatryan:2014iha,Aad:2015xua,Khachatryan:2015mma,Khachatryan:2016ctc,Aaboud:2018puo,Sirunyan:2019twz,CMS:2022ley,CMS:2024eka}
broadly consistent with the SM predictions, as measured by the CMS and ATLAS Collaborations at the LHC.
To date, no significant deviation from the SM predictions regarding the Higgs boson properties or
its interactions has been observed, but several areas within the Higgs sector remain to be explored.

The study of the evolution of $\PH$-mediated and continuum diboson production as a function
of the diboson invariant mass, when the Higgs boson is produced both on- and off-shell, enables
the extraction of \GH from the mass dependence of the Higgs boson propagator~\cite{CMS:2024eka,
CMS:2022ley, Caola:2013yja, Kauer:2012hd, Campbell:2013una}.
The on- and off-shell cross-sections for a given production mode $j$ for vector-boson decays are proportional to 
\begin{equation}
    \sigma_j^\text{on} \propto \frac{g_p^2 g_d^2}{\GH}, \qquad
    \sigma_j^\text{off} \propto g_p^2 g_d^2,
    \label{eq:offshell_xsec}
\end{equation}
where $g_p$ and $g_d$ represent the set of couplings associated with Higgs boson production and decay, respectively.
A simultaneous fit to both regions then provides constraints on \GH. This so-called ``off-shell method'' exploits the 
interference between the \PH-mediated amplitude and two-vector-boson production, since evidence 
for off-shell Higgs boson production appears through the negative interference with the gluon-fusion and electroweak (EW)
backgrounds. The on-shell region is defined as the region of the peak around 125\GeV.
The region used to analyze off-shell Higgs boson production begins above $2m_\PZ$;
for the reasons given in Section~\ref{sec:Discriminants}, it is taken to start at $220\GeV$.

This study builds upon the analysis presented in Ref.~\cite{CMS:2024eka}, in which \GH is measured 
using production in the off-shell region, to explore various beyond-SM (BSM) hypotheses.
While most studies are performed in the $\PH\to\PZ\PZ\to4\ell$ channel, we also carry out a combined analysis of the 
$\PH\to\PZ\PZ\to4\ell$, $2\ell2\nu$, and $\PH\to\PW\PW\to2\ell2\nu$ channels to improve sensitivity to the Higgs boson width.

We consider modifications to the following Higgs boson (self-)interactions.
The effective Lagrangian describing the trilinear ($\lambda_3$) and quartic ($\lambda_4$) Higgs boson self-interactions is
\begin{equation}
\label{eq:kappa_lambda}
{\mathcal{L}}_\text{self} = -\lambda_3 v h^3 - \frac{\lambda_4}{4} h^4,
\end{equation}
where $h$ denotes the physical Higgs field and $v\simeq246\GeV$ is the vacuum expectation value.
We define the Higgs boson trilinear self-coupling modifier $\kappa_{\lambda}={\lambda_3}/{\lambda_3^\mathrm{SM}}$,
where the SM value is $\lambda_3^\mathrm{SM}=\lambda_4^\mathrm{SM}=m_\PH^2/(2v^2)\simeq 0.13$.
The \CP-even and \CP-odd Yukawa couplings $y_{\PQq}$ and $\tilde{y}_{\PQq}$ of a quark $\PQq$
are incorporated into the effective Lagrangian that describes its interaction with the Higgs boson as
\begin{equation}
\label{eq:LagrHqq}
{\mathcal{L}}_{\PH\qqbar}=-\overline{\psi}_{\PQq}(y_{\PQq} + i \tilde{y}_{\PQq} \gamma^5 ){\psi}_{\PQq}h,
\end{equation}
where $\overline{\psi}_{\PQq}$ and ${\psi}_{\PQq}$ denote the Dirac spinors for a quark, $\PQq$,
which can be $\PQu$, $\PQd$, $\PQs$, $\PQc$, $\PQb$, $\PQt$, or $\PQQ$. We include
a BSM heavy quark $\PQQ$, which models any heavy BSM contributions to the gluon-fusion (\ggH) loop.
We define $\kappa_{\PQq}=y_{\PQq}v/m_{\PQq}$ and $\tilde{\kappa}_{\PQq}=\tilde{y}_{\PQq}v/m_{\PQq}$,
which are useful since $\kappa_{\PQq}^\mathrm{SM}=1$ for the SM quarks.
We also define $\barkappaq=y_{\PQq}v/m_{\PQb}$, where $m_{\PQb}$ is the mass of the bottom-quark,
which is particularly useful for comparing the hierarchy
of the Yukawa couplings of light quarks, with respect to $\barkappab^\mathrm{SM}=1$.
In these calculations, the quark masses are evaluated at the scale $\mu_\text{QCD}=125\GeV$~\cite{PhysRevD103016010}.

In recent measurements, potential \CP-even contributions
from heavy quarks in the \ggH loop were included to assess the stability of the off-shell method
concerning BSM effects~\cite{CMS:2024eka}. This study aims to expand upon these analyses by incorporating
both \CP-odd and \CP-even effects from heavy quarks, such as the top or BSM fourth-generation
quarks, as well as modified couplings
to light quarks, facilitating a more model-independent understanding of \GH constraints.
Additionally, we allow for custodial symmetry breaking by not imposing equal couplings of the 
Higgs boson to the $\PZ$ and $\PW$ bosons, which are modified by $\kappa_\PZ$ and $\kappa_\PW$, respectively.

While previous investigations have explored the couplings of the Higgs boson to heavy quarks in off-shell \Hell
decays~\cite{CMS:2024eka}, the couplings to light quarks were not investigated.
Constraints on the light-quark Yukawa couplings have been studied in on-shell \Hell decays~\cite{CMS:2025xkn}.
In those studies, custodial symmetry and the constraint $\abs{\kappa_\PZ} \leq 1$ were imposed due to 
the parameterization 
of the cross section related to \GH and $\kappa_\PZ$, which become degenerate for $\abs{\kappa_\PZ} > 1$.
Although sensitivity to light-quark Yukawa couplings is significantly diminished in the off-shell region, the
rate of off-shell production remains independent of \GH, as seen in Eq.~(\ref{eq:offshell_xsec}),
breaking the degeneracy between \GH and $\kappa_\PZ$ in determining the signal rate.

There are two primary approaches to constraining the Higgs boson self-coupling: a direct method via
$\PH\PH$ production and an indirect method through loop corrections in single Higgs boson production.
We perform a measurement of the $\kappa_{\lambda}$
using the latter approach within the off-shell regime~\cite{Haisch:2021hvy};
in this regime, modifications of $\kappa_{\lambda}$ uniquely distort the \mell distribution, 
providing sensitivity in a previously unexplored region of phase space.
The $\kappa_{\lambda}$ corrections are incorporated for both \ggH~\cite{Haisch:2021hvy} and
EW production processes. Although this measurement is expected to have lower sensitivity 
than $\PH\PH$ production~\cite{2025139210}, it provides a complementary means
of constraining the Higgs boson self-coupling.

One possible solution to the hierarchy problem, wherein the $m_\PH$ receives large quantum corrections,
is the composite Higgs boson model, which posits that the Higgs boson is not elementary but rather a bound state of
more fundamental constituents.  An analogy can be drawn with hadronic states, such as the composite spin-zero pion.
In this study, we establish limits on one such composite Higgs boson model utilizing
the off-shell region. For a review of composite Higgs models, which have yet to be excluded at higher energy
scales, see Ref.~\cite{Witzel:2019jbe}.

Composite Higgs models introduce new strong dynamics at the TeV mass scale, and Higgs boson interactions may
exhibit a momentum-dependent form-factor at the BSM physics scale $\Lambda_\PH$~\cite{Han:2023krp},
analogous to the phenomenological form-factors used to describe hadronic structure.
Essentially, the form-factor measures the deviation of a composite object from point-like behavior in scattering processes.
The form-factor is given by
 \begin{equation}
     F(q^2) =
     \left[
     \frac{1}{1 + \abs{ q^2 } / \Lambda_\PH^2}
     \right]^n,
     \label{eq:formfact-composite}
 \end{equation}
where $\Lambda_\PH$ represents
the compositeness scale and $q^2$ denotes the momentum transfer of the Higgs boson.
The choice $n=1$ or $n=2$ can be viewed as representative monopole- or dipole-like behavior. 
We adopt $n=1$ as the standard benchmark when the form-factor is applied to both production
and decay vertices, or equivalently $n=2$ when it is applied to a single production or decay vertex.
Larger values of $n$ would amplify the form-factor effect and result in stronger constraints.

It is worth noting that the value of $\Lambda_\PH$ is associated with the characteristic size $d$ of a composite object,
following the relationship
\begin{equation}
    d= \frac{\hbar c}{\Lambda_\PH}.
        \label{eq:size-composite}
\end{equation}
The off-shell regime of Higgs boson production serves as an optimal testing ground for composite Higgs models,
as the momentum transfer $q^2$ of the Higgs boson ranges from approximately $m_\PH^2$ up to $(1\,\TeV)^2$
and beyond. Such a test cannot be performed using on-shell Higgs boson production, unless we restrict
our tests to $\Lambda_\PH\sim m_\PH$.

The rest of the paper is organized as follows.
Sections~\ref{sec:Detector}--\ref{sec:Discriminants} present the CMS detector and the data samples used in the
analysis, along with the event reconstruction, selection, and kinematic discriminants defining the off-shell signal region.
The likelihood model, simulated samples, and systematic uncertainties used in the fit are then described in
Section~\ref{sec:SignalModelling}. Next, the off-shell measurement is interpreted in terms of Higgs boson substructure
in Section~\ref{sec:H_struct}, heavy-particle effects in the gluon-fusion loop in Section~\ref{sec:H_heavy},
light-quark Yukawa couplings in Section~\ref{sec:H_Yukawa}, and the Higgs boson self-coupling in
Section~\ref{sec:kappa_lambda}. Constraints on the total Higgs boson width under increasingly general
BSM hypotheses are extracted in Section~\ref{sec:width}, and multiple on- and off-shell channels are finally combined
to obtain the most precise width determination and the corresponding off-shell signal-strength measurement
in Section~\ref{sec:comb}. Tabulated results are provided in the HEPData record for this analysis in Ref.~\cite{hepdata}.

\section{The CMS detector}\label{sec:Detector}
The central feature of the CMS detector is a superconducting solenoid of 6\unit{m} internal diameter,
providing a magnetic field of 3.8\unit{T}.
Within the solenoid volume are a silicon pixel and strip tracker,
a lead tungstate crystal electromagnetic calorimeter (ECAL), and a brass and scintillator
hadron calorimeter (HCAL), each composed of a barrel and two endcap sections.
Forward calorimeters extend the pseudorapidity ($\eta$) coverage provided by the barrel and endcap detectors.
Muons are reconstructed in gas-ionization detectors embedded in the steel flux-return yoke outside the solenoid.
More detailed descriptions of the CMS detector, together with a definition of the coordinate system used
and the relevant kinematic variables, can be found in Refs.~\cite{CMS:2008xjf,CMS:2023gfb}.

Events of interest are selected using a two-tiered trigger system~\cite{Khachatryan:2016bia}. The first level, composed of custom hardware
processors, uses information from the calorimeters and muon detectors to select events at a rate of around 100\unit{kHz}
within a fixed latency of about 4\mus~\cite{Sirunyan:2020zal}. The second level, known as the high-level trigger,
consists of a farm of processors running a version of the full event reconstruction software optimized for fast processing,
and reduces the event rate to around 1\unit{kHz} before data storage~\cite{Hayrapetyan_2024}.

The primary vertex is taken to be the vertex corresponding to the hardest scattering in the event,
evaluated using tracking information alone, as described in Section 9.4.1 of Ref.~\cite{CMS-TDR-15-02}.

The electron momentum is estimated by combining the energy measurement in the ECAL with the momentum
measurement in the silicon tracker. The transverse momentum (\pt) resolution ranges from 1.6 to 5\% for electrons with
$\pt \approx 45\GeV$ from $\PZ \to \Pe \Pe$ decays.
It is generally better in the barrel region than in the endcaps, and also depends on the
bremsstrahlung energy emitted by an electron as it traverses the material in front of the
ECAL~\cite{eleReco,ScaleSmear2}.

Muons are measured in the range of $\abs{\eta} < 2.4$, with detection planes made using three
technologies: drift tubes, cathode strip chambers, and resistive-plate chambers.
Matching muons to tracks measured in the silicon tracker results in a relative \pt
resolution of 1\% in the barrel and 3\% in the endcaps, for muons with $\pt<100\GeV$, and of better than 7\% in the barrel for muons with
$\pt<1\TeV$~\cite{MuReco}.

Jets are clustered from particle-flow candidates using the anti-\kt algorithm~\cite{Cacciari:2008gp,Cacciari:2011ma}
with a distance parameter of 0.4. The jet momentum is defined as the
vector sum of all particle momenta in a jet and is found from simulation to be, on average,
within 5--10\% of the true momentum over the entire \pt spectrum and detector acceptance \cite{Khachatryan_2017}.

\section{Data and simulated samples}\label{sec:Data}

The Run~2 data sample of proton-proton collisions used in this analysis corresponds
to integrated luminosities of $36.3$, $41.5$, and $59.8\fbinv$ collected in 2016, 2017, and 2018, respectively, 
for a total of $138\fbinv$ recorded at a center-of-mass energy of 13\TeV.
Events are selected online using the same set of triggers as adopted in Ref.~\cite{Sirunyan:2021rug}.
They require the presence of at least one lepton (either muon or electron),
or up to three leptons with relaxed \pt conditions.
The trigger efficiency relative to the offline selection discussed in Section~\ref{sec:EventSelection} is found to be larger than 99\%,
measured in data using 4-lepton events collected by the single-lepton triggers.
It agrees with the expectation from simulation at the per mille level.

Monte Carlo (MC) simulations are used to model both signal processes, which involve the Higgs boson,
and background processes in proton-proton ($\Pp\Pp$) interactions and their reconstruction in the CMS detector.
All simulated samples are processed with \PYTHIA~8.320~\cite{Sjostrand:2014zea} to simulate
multiparton interactions, parton showering, and hadronization.
The CP5 tune~\cite{Sirunyan:2019dfx}
is used for the simulation of the entire data-taking period.
The NNPDF~3.1 parton distribution functions~\cite{Ball:2011uy} are used for all simulated samples.
Simulated events include the contribution from additional $\Pp\Pp$ interactions within the same or adjacent
bunch crossings (pileup) and are weighted to reproduce the pileup distribution observed in data.
The simulated events are further processed through a dedicated simulation of the CMS detector based
on \GEANTfour~\cite{Agostinelli2003250}.

\begin{figure*}[!bht]
    \centering
    \includegraphics[width=0.4\textwidth]{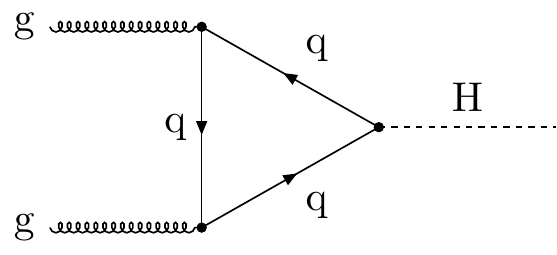}
    \includegraphics[width=0.4\textwidth]{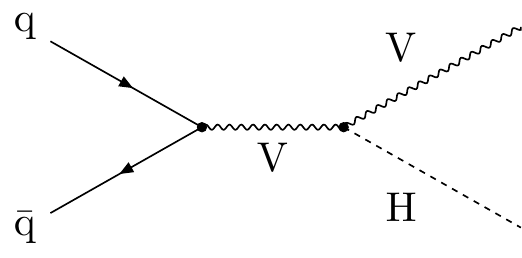} \\
    \includegraphics[width=0.4\textwidth]{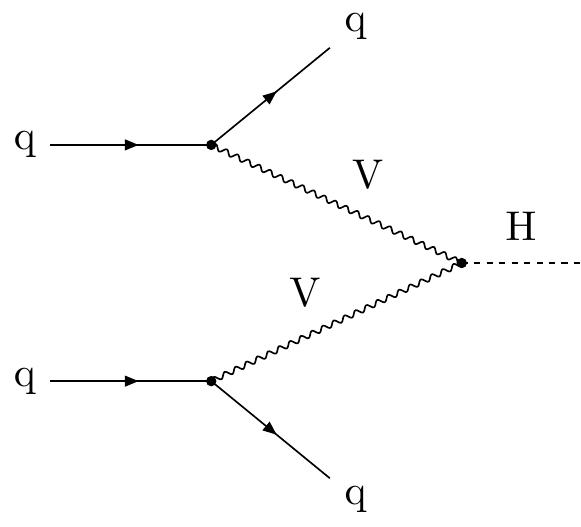}
    \includegraphics[width=0.4\textwidth]{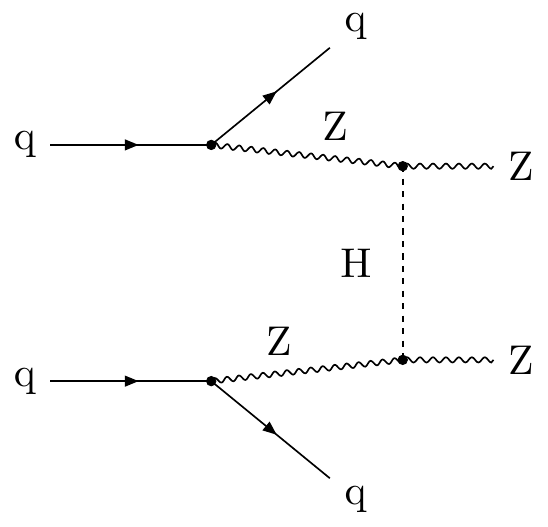}
    \caption{
    Leading-order Feynman diagrams for the Higgs boson production processes considered in the \offshell region:
    \ggH (upper left), $\PV\PH$ associated production (upper right),
    $s$-channel VBF (lower left), and $t$-channel VBF (lower right).
      }\label{fig:prod_modes}
\end{figure*}

\begin{figure}[!thb]
    \centering
    \includegraphics[width=0.45\textwidth]{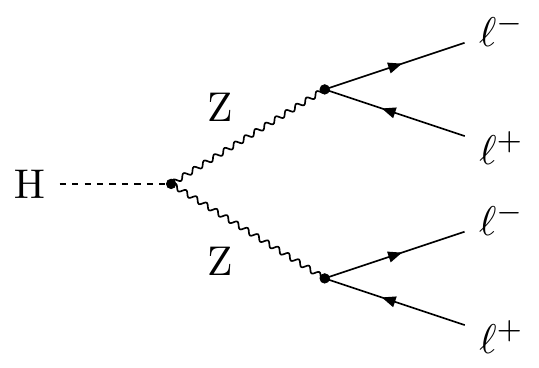}
    \caption{
    The leading-order Feynman diagram illustrating the $\PH\to\PZ\PZ\to4\Pell$ decay process.
     }
    \label{fig:decay_mode}
\end{figure}

The samples modeling off-shell Higgs boson production include the interference between diagrams with and
without the Higgs boson in the propagator. Both \ggH and EW Higgs boson production
(the latter including vector-boson fusion (VBF) and associated production ($\PV\PH$) modes, where $\PV$ denotes the $\PZ$ and $\PW$ bosons)
are considered in the off-shell region and are illustrated in Fig.~\ref{fig:prod_modes}.
As shown in Fig.~\ref{fig:prod_modes}, VBF production via the $t$~channel is only possible
with \PZ bosons.
For studies of enhanced light-quark Yukawa couplings, the direct quark-antiquark
annihilation channel producing the Higgs boson is also included.
In all cases, we consider the decay process $\PH\to\PZ\PZ\to4\Pell$, where $\Pell=\PGm,\Pe$, as illustrated in Fig.~\ref{fig:decay_mode}.

The $\Pg\Pg\to\ZZ$ / $\PZ\gamma^{\ast}\to4\ell$ process is simulated with
\mcfm~7.0.1~\cite{MCFM,Campbell:2011bn,Campbell:2013una,Campbell:2015vwa} interfaced with the \jhugen~7.5.7~
generator framework. The \jhugen 
framework~\cite{Gao:2010qx,Bolognesi:2012mm,Anderson:2013afp,Gritsan:2016hjl,Gritsan:2020pib,Davis:2021tiv}
is also used to simulate the EW production
components: vector-boson scattering, triple-gauge boson backgrounds (VVV), the EW signal
(VBF and $\VH$), and their interference. The framework, furthermore, allows the implementation of anomalous
interactions of the Higgs boson or the gauge bosons in both \ggH and EW production modes.
The SM simulation of EW processes is validated against \textsc{phantom}~1.3~\cite{Ballestrero:2007}. 
The Matrix-Element Likelihood Approach (\mela) package~\cite{Gao:2010qx,Bolognesi:2012mm,Anderson:2013afp,Gritsan:2016hjl,Gritsan:2020pib} provides
a library of matrix elements from \jhugen (signal) and  \mcfm (background and interference), which is used to
reweight generated off-shell samples. Samples are produced for a variety of anomalous coupling hypotheses
together with the SM and then reweighted with \mela to cover the range of BSM scenarios studied here.
All off-shell production simulation and matrix element calculations are performed at leading order (LO), 
with higher-order effects accounted for via K-factors, as described below.

Wide resonances produced via \ggH, VBF, or $\PV\PH$ with masses from 115 to 3000\,\GeV 
are simulated with the next-to-leading-order (NLO) quantum chromodynamics (QCD) event generator 
\POWHEG~2.0~\cite{Frixione:2007vw,Bagnaschi:2011tu,Nason:2009ai,Luisoni:2013kna,Hartanto:2015uka,PowhegBox_1,PowhegBox_2}, 
while \jhugen models their decays to four leptons.
These samples are employed to study and validate the associated jet activity
across the broad off-shell region, which is crucial for event categorization. Events from the \POWHEG + \jhugen simulation
are reweighted with \mela to obtain alternative off-shell Higgs boson distributions, including interference with the background.
This complements the direct off-shell simulation and provides useful cross checks. Both approaches rely on the same
\mela/\jhugen matrix elements, and starting from NLO \POWHEG production matched to a parton shower improves
the modeling of accompanying jet activity compared with LO off-shell samples.

In the \ggH process, the factorization and renormalization scales are chosen to run as $\mell/2$.
Higher-order QCD corrections are evaluated by computing signal cross sections at LO, NLO, and next-to-NLO (NNLO)
using \mcfm and \textsc{HNNLO}~2~\cite{Catani:2007vq,Grazzini:2008tf,Grazzini:2013mca} in the narrow-width approximation
over the relevant mass range. The NNLO-to-LO ratios (K-factors) are used to reweight the $\mell$ distributions from
the LO \MCFM and \jhugen simulations~\cite{deFlorian:2016spz}. A uniform-factor of 1.10 is applied across
the full $\mell$ range to normalize the \ggH Higgs boson production cross section to the next-to-NNLO (N${3}$LO) prediction
near $\mell\approx 125\GeV$~\cite{deFlorian:2016spz}. The $\POWHEG+\jhugen$ \ggH samples
are reweighted using NNLO-to-NLO K-factors to improve their modeling of higher-order effects.

While the NNLO-to-LO K-factor calculation can be directly applied to the signal cross section,
its use for the \ggZZ background and the signal-background interference is only approximate.
Approximate NLO calculations for the background and interference are
available~\cite{Caola:2015psa,Melnikov:2015laa,Campbell:2016ivq,Caola:2016trd}.
Replacing the values from the simulation programs with these NLO results increases
the K-factor for the interference near the $m_{4\ell}$ region corresponding  to the kinematic
threshold for two \onshell $\PZ$ bosons. Nevertheless, in the mass range relevant to this
analysis, $\mell > 220\GeV$, the NLO-to-LO K-factors for both the background and the
interference remain consistent with that of the signal within about 10\%. We therefore apply
 the same NNLO-to-LO K-factor and uniform N$^3\text{LO}$ correction,
 both derived for the signal, and including their associated uncertainties,
 to the background and interference contributions. An additional 10\% uncertainty is assigned
 to the background to account for this approximation, while the interference is assigned
 an uncertainty equal to the square root of this value.

The $\PQq\PAQq\to\PH\to4\Pell$ signal is generated at LO using \MGvATNLO~v2.4.2~\cite{Alwall:2014hca}
over the full $m_{4\ell}$ range. An $m_{4\ell}$-dependent K-factor, evaluated in the
\offshell region with \textsc{SusHi}~\cite{Harlander:2012pb}, is applied to 
approximate NNLO QCD accuracy.
The calculations use a factorization scale of $m_{4\ell}/4$ and a renormalization scale of $m_{4\ell}$,
in line with Ref.~\cite{CMS:2025xkn}, which were found to provide the most stable predictions.
The interference with the $\PQq\PAQq \to 4\ell$ background is neglected, since it vanishes at LO
because of the different polarizations of the initial-state partons.

The $\PQq\PAQq\to4\Pell$ background process is simulated at NLO in QCD and LO in the EW theory
using \POWHEG. The fully differential cross section for this process has been calculated at NNLO in
QCD~\cite{Grazzini:2015hta}, and the resulting NNLO-to-NLO K-factor, parameterized as a function
of $m_{4\ell}$, is applied to the \POWHEG prediction. This correction is approximately 1.1 at
$m_{4\ell} = 125\GeV$ and varies between 1.0--1.2 for $m_{4\ell} < 500\GeV$.
An EW NLO-to-LO K-factor~\cite{Bierweiler:2013dja} is additionally applied for events with
two \onshell $\PZ$ bosons. This correction depends on both $m_{4\ell}$ and the $t$-channel
momentum transfer squared, decreasing from unity near $125\GeV$ to approximately 0.9
in the region $m_{4\ell} > 500\GeV$. The uncertainty associated with this correction constitutes
the dominant systematic uncertainty in the off-shell \GH measurement and is implemented
as a function of \mell.

\section{Event reconstruction and selection}\label{sec:EventSelection}
The event reconstruction is based on the particle-flow (PF) algorithm~\cite{Sirunyan:2017ulk}.
This algorithm aims to reconstruct and identify each individual particle in an event, with an optimized
combination of information from the various elements of the CMS detector. The energy of photons is obtained
from the ECAL measurement. The energy of electrons is determined from a combination of the electron momentum
at the primary vertex as determined by the tracker, the energy of the corresponding ECAL cluster, and the
energy sum of all bremsstrahlung photons spatially compatible with originating from the electron track.
The energy of muons is obtained from the curvature of the corresponding track.
The energy of charged hadrons is determined from a combination of their momentum measured in the tracker and
the matching ECAL and HCAL energy deposits, corrected for the response function of the calorimeters to hadronic
showers. Finally, the energy of neutral hadrons is obtained from the corresponding corrected ECAL and
HCAL energies.

Electrons with $\pt>7\GeV$ are reconstructed within the geometrical acceptance, corresponding to the
region of $\abs{\eta}<2.5$~\cite{eleReco}. They are identified using a multivariate
discriminant, which includes observables sensitive to the emission of bremsstrahlung along the
electron trajectory, the geometrical and momentum-energy matching between the electron trajectory
and the associated cluster in the ECAL, the shape of the electromagnetic shower in the ECAL, and
variables that discriminate against electrons originating from photon conversions. The isolation
parameter sums for electrons, defined similarly as for muons, are included in the multivariate
discriminant. This information is used to discriminate between prompt leptons from \PZ boson decays
and those arising from EW decays of hadrons within jets. The package \textsc{XGBoost}~\cite{xgboost}
is used to train and optimize this multivariate discriminant. The training is performed with simulated
events that are not used at any other stage of the analysis. Separate trainings are performed for the
three different data-taking periods~\cite{ElectronBDT}. Electrons from photon conversions and muons
from nonprompt decays of hadrons are rejected if the impact parameter significance in three dimensions,
defined as the ratio of the impact parameter with respect to the primary vertex to its uncertainty, is greater than four.

Muons with $\pt > 5\GeV$ are reconstructed within the geometrical acceptance, corresponding to the region
$\abs{\eta}< 2.4$, by combining information from the silicon tracker and the muon systems~\cite{MuReco}.
The muons are selected among the reconstructed muon track candidates by applying quality requirements on the
track in both the muon and silicon tracker systems, and demanding small energy deposits in the calorimeters.
A relative muon isolation variable is defined as
\begin{widetext}    
\begin{equation} \label{eqn:pfiso}
  {\mathcal{I}}^{\PGm} \equiv \Big( \sum \pt^\text{charged} + \max\big[ 0, \sum \pt^\text{neutral} +
    \sum \pt^{\PGg} - \pt^{\mathrm{PU}} \big] \Big) / \pt^{\PGm},
\end{equation}
\end{widetext}
where $\pt^{\PGm}$ is the muon transverse momentum, and $\pt^{\text{charged}}$, $\pt^\text{neutral}$, and $\pt^\PGg$
are the transverse momenta of charged hadrons, neutral hadrons, and photons, respectively, within a cone radius
of $\Delta R=0.3$ around the muon direction at the primary vertex.\\
Here, $\Delta R(i,j) \equiv \sqrt{\smash[b]{(\eta^i-\eta^j)^{2} + (\phi^i-\phi^j)^{2}}}$, where $\phi$
is the azimuthal angle in radians, $\eta$ the pseudorapidity, and in this case $i$ refers 
to the hadron or photon, and $j$ to the muon.
Since the isolation variable is particularly sensitive to energy deposits from pileup interactions, a
contribution $\pt^{\mathrm{PU}}$ from pileup is subtracted from the isolation parameter, as shown in
Eq.~(\ref{eqn:pfiso}).
It is defined as 0.5 times the \pt sum of all charged hadrons $i$ not originating from the primary
vertex $\pt^{\mathrm{PU}} = 0.5\sum_{i}\pt^{i,\mathrm{PU}}$, where the 0.5 factor accounts for the
different fraction of charged and neutral hadrons~\cite{PUmitigationCMS}. A requirement of
${\mathcal{I}^{\PGm}} < 0.35$ is placed on each muon in the event.

Photons from final-state radiation are reconstructed using the PF algorithm~\cite{Sirunyan:2017ulk}. The isolation sum $\mathcal{I}^{\PGg}$ is
constructed as the sum of photon, neutral, and charged isolation components within a cone of $\Delta R < 0.3$, and is
normalized with the photon \pt.
Isolated photons with $\pt >2\GeV$, $\abs{\eta} <2.4$, and $\mathcal{I}^{\PGg} < 1.8$,
are associated with the closest lepton (either muon or electron) in the event.
Photons that do not satisfy the requirements $\Delta R(\PGg,\Pell)/(\pt^{\PGg})^2<0.012\GeV^{-2}$
and $\Delta R(\PGg,\Pell) < 0.5$ are discarded.
If more than one photon candidate fulfils the above conditions, the one with the lowest value of
$\Delta R(\PGg,\Pell)/(\pt^{\PGg})^2$ with respect to the given lepton is retained.
Photons passing the above criteria are excluded from the computation of the relative isolation parameter.

The reconstruction and selection efficiencies for prompt leptons in both data and simulation have been
estimated using a tag-and-probe technique~\cite{CMS:2011aa} based on samples of $\PZ$ boson events.
The ratio of the efficiencies measured in data and simulation is used to rescale the yields of selected
events in the simulated samples.
In addition, $\PZ$ boson events have been used to calibrate the momentum scale and resolution of
electrons and muons in bins of different kinematic variables~\cite{ScaleSmear2, Roch2}.

In the selection of Higgs boson candidates,
four prompt and isolated leptons are required following the prescription above.
One of the leptons must have $\pt>20\GeV$ and at least one of the remaining leptons must satisfy $\pt>10\GeV$.
The \PZ boson candidates are constructed from $\Pep\Pem$ or $\PGmp\PGmm$ pairs whose invariant mass
is in the range 12--120\GeV.
The dilepton pairs are then combined to form the Higgs boson candidate.
In the following, $\PZ_1$ denotes the pair with the mass closest to the nominal $\PZ$ boson mass,
while $\PZ_2$ refers to the remaining one. Four possible combinations are considered and treated separately:
4\PGm, 4\Pe, $2\Pe2\PGm$, and $2\PGm2\Pe$, where the mixed-flavor final states are separated based
on the decay of $\PZ_{1}$.
The four possible combinations have different \mell resolutions (largely driven by
whether the $\PZ_{1}$ is formed from 2$\PGm$ or 2$\Pe$)
and different amounts of reducible background (largely driven by whether the $\PZ_{2}$ decays to 2\PGm or 2\Pe).
None of the above differences between the flavor combinations affects the \offshell Higgs boson analysis
and, therefore, all flavor combinations are combined in that measurement.
Signal candidates must satisfy $m_{4\Pell} > 70\GeV$.
If more than one Higgs boson candidate can be formed in the event, the one with the highest value of the
kinematic discriminant $\mathcal{D}_\text{bkg}^\text{kin}$, defined in Section~\ref{sec:Discriminants},
is retained, unless these candidates consist of the same four leptons.
In this case, the candidate with the $\PZ_1$ invariant mass closest to the nominal \PZ boson mass is retained.

In the \offshell analysis, events are further categorized based on the jets associated with the Higgs boson candidate.
Jets must satisfy $\pt>30\GeV$ and $\abs{\eta}<4.7$, and must be separated
from all selected lepton candidates and any selected final-state radiation photons by demanding
$\Delta R(\Pell/\PGg,{\text{jet}})>0.4$.
Jets originating from the hadronization of \PQb quarks are identified using the \textsc{DeepCSV}
algorithm~\cite{Sirunyan:2017ezt}, which combines information
on impact parameter significance, the secondary vertex, and jet kinematic variables.
The use of this identification is described in Section~\ref{sec:Discriminants}.

\section{Kinematic observables and event categorization}\label{sec:Discriminants}
The principal kinematic variable used in the study of the $\PH\to\PZ\PZ\to 4\Pell$ channel is the
\mell. In the off-shell region, the $m_{4\Pell}$ spectra of the signal and background
are not sharply distinct, and their interference can produce a net deficit when signal and interference are combined.
Many of the effects examined in this paper are sensitive to the shape of the signal $m_{4\Pell}$ distribution
and on how it is modified by interference with the background. Therefore, $m_{4\Pell}$ is treated as a primary observable.
The distribution is divided into 21 mass bins in the region $m_{4\Pell}>220\GeV$. This lower bound was chosen
to avoid the complex background behavior at the $2m_{\PZ}$ threshold while retaining a substantial portion
of potential Higgs boson signal. 

In the on-shell region, which for Ref.~\cite{CMS:2025xkn} is between 105 and 140\GeV,
$m_{4\Pell}$ primarily serves to separate signal from background, since the signal
produces a pronounced peak at $m_{\PH}$ and interference with background is negligible.
Because the detector resolution for $m_{4\Pell}$ is of order $1\GeV$, far exceeding the
Higgs boson intrinsic width $\Gamma_{\PH}=4.1\MeV$, this observable cannot probe the Higgs boson
propagator beyond the pole mass $m_{\PH}=125.38\GeV$, which is used to be consistent with Ref.~\cite{CMS:2025xkn}.
The primary objective in the on-shell region is to normalize the Higgs boson production cross section.
This is achieved using a single signal strength modifier, $\mu_{\text{on}}$, which parameterizes
the on-shell signal and is determined from a fit to the discriminant $\mathcal{D}_{\text{sig}}$
computed via Eq.~(\ref{eq:melaD}), discussed below. In that calculation, the signal and background
probabilities appearing in the numerator and denominator incorporate the reconstructed $m_{4\Pell}$ probability.

\subsection{Kinematic discriminants}

To achieve full sensitivity to a signal process with a sizable background interference component,
such as both \ggH and EW production of the Higgs boson in the off-shell region,
two discriminants are used: $\mathcal{D}_{\text{sig}}$ and $\mathcal{D}_{\text{int}}$, which are defined as
\begin{equation}
\mathcal{D}_\text{sig}\left(\boldsymbol{\Omega}\right) = \frac{\mathcal{P}_\text{sig}\left(\boldsymbol{\Omega}\right) }
        {\mathcal{P}_\text{sig}\left(\boldsymbol{\Omega}\right) +\mathcal{P}_\text{alt}\left(\boldsymbol{\Omega}\right) },
\label{eq:melaD}
\end{equation}
\begin{equation}
\mathcal{D}_\text{int}\left(\boldsymbol{\Omega}\right) =
\frac{\mathcal{P}_\text{int}\left(\boldsymbol{\Omega}\right) }
{2 \ \sqrt{{\mathcal{P}_\text{sig}\left(\boldsymbol{\Omega}\right) \ \mathcal{P}_\text{alt}\left(\boldsymbol{\Omega}\right) }}},
\label{eq:melaDint}
\end{equation}
where the labels ``sig'' and ``alt'' on each probability $\mathcal{P}$ denote the 
signal model and an alternative model (which may represent a different \PH production
mechanism or a background, as appropriate), while  ``int'' denotes the interference between the two contributions,
and $\boldsymbol{\Omega}$ the set of up to 13 kinematic observables containing the
production and decay angles, alongside the masses of any intermediate states,
that describe the event~\cite{Gritsan:2020pib}.

These discriminants, used in combination, provide the optimal separation between signal and background
in the presence of their interference, irrespective of the relative magnitudes of the contributing amplitudes.
The kinematic observables in $\boldsymbol{\Omega}$ are too numerous to handle directly in a single analysis. 
Therefore, the \mela is employed
to reduce the dimensionality: it compresses the relevant kinematic information into a likelihood 
quantity $\mathcal{P}$ using analytically
derived matrix element calculations~\cite{Gao:2010qx,Bolognesi:2012mm,Anderson:2013afp,Gritsan:2016hjl,Gritsan:2020pib},
as illustrated in Eqs.~(\ref{eq:melaD}) and~(\ref{eq:melaDint}).

\subsection{Event categorization}\label{sec:categorization}

Reconstructed four-lepton events, together with their associated jets, can be characterized by
observables that exploit the kinematic features of Higgs boson production and decay,
for example, the additional jet activity in VBF. These kinematic differences can be used to divide events into categories. In this analysis we define three such categories: VBF-tagged, VH-tagged,
and untagged, corresponding approximately to VBF-like, $\PV\PH$-like, and $\Pg\Pg$- or
$\PQq\PAQq$-initiated topologies, respectively. Within each category, the \mell distribution
is binned together with a discriminant to form the analysis observables.

We apply a set of kinematic discriminants together with other selection criteria to classify the off-shell events.
These discriminants are variants of $\mathcal{D}_{\text{sig}}$ from Eq.~(\ref{eq:melaD}), specifically constructed
to separate the signal production mechanisms \VBF, \WH, \ZH, and \ggH.
Distinguishing four production modes requires three independent observables, denoted
$\mathcal{D}_{\text{2jet}}^{\VBF}$, $\mathcal{D}_{\text{2jet}}^{\ZH}$,
and $\mathcal{D}_{\text{2jet}}^{\WH}$. The labels indicate the dijet topology (``2jet'')
and the targeted production mechanism. Further details are given in
Refs.~\cite{Sirunyan:2017exp,Sirunyan:2017tqd,Sirunyan:2019twz,Sirunyan:2021rug}.

The $\mathcal{D}_{\text{2jet}}$ discriminants are built according to Eq.~(\ref{eq:melaD}),
where $\mathcal{P}_{\text{sig}}$ is the signal probability for \VBF (or $\WH$/$\ZH$) production
in the VBF-tagged (or \VH-tagged) category, and $\mathcal{P}_{\text{alt}}$ corresponds to \PH
production via \ggH in association with two jets. If more than two jets satisfy the selection,
the two jets with the highest \pt are used in the matrix element evaluation.
The $\mathcal{D}_{\text{2jet}}$ discriminants distinguish the targeted production mode
in each category from \ggH production using only the kinematic information of the Higgs boson
and the two associated jets. Additional jet-related requirements involve jets identified as
originating from \PQb quark hadronization (\PQb-tagged jets)~\cite{Sirunyan:2021rug}.
The number of \PQb-tagged jets is restricted because events with many \PQb-tagged jets
resemble top-quark backgrounds. The categories are then defined by applying the following
sequence of selection criteria in the following order:

\begin{itemize}
\item[--]  The {$\VBF$}-tagged category requires exactly four leptons. In addition, there must be
either two or three jets of which at most one is \PQb-tagged,
or at least four jets and no \PQb-tagged jets.
Finally, $\mathcal{D}_\text{2jet}^{\VBF} >0.5$ is required~\cite{Sirunyan:2021rug}.
\item[--]  The {$\VH$-tagged} category requires exactly four leptons. In addition, there must be
either two or three jets, or at least four jets and no \PQb-tagged jets.
Finally, we demand $\max(\mathcal{D}_\text{2jet}^{\WH},\mathcal{D}_\text{2jet}^{\ZH} )>0.5$~\cite{Sirunyan:2021rug}.
\item[--]  The {untagged} category consists of the remaining four-lepton events which pass selection.
\end{itemize}

\subsection{Kinematic variables used in the fit}

To be sensitive to the Higgs boson in the off-shell region, one must both distinguish signal-like events from the
large $\PQq\PAQq \to 4\Pell$ background and preserve sensitivity to interference effects.

For each category we construct two discriminants. One discriminant, $\mathcal{D}_{\text{sig}}$, is optimized
to separate the Higgs boson signal from the dominant $\PQq\PAQq\to 4\Pell$ background
and is defined as in Eq.~(\ref{eq:melaD}).
The second discriminant, $\mathcal{D}_{\text{int}}$, is optimized to isolate the interference between signal
and background for the relevant production mode in that category, \ggH in the untagged category,
and EW production in the \VBF- and \VH-tagged categories, and is defined as in Eq.~(\ref{eq:melaDint}).
In the untagged category only the \PH decay information is used, and the discriminants
are denoted $\mathcal{D}_{\text{sig}}^{\text{dec}}$ and $\mathcal{D}_{\text{int}}^{\text{dec}}$, respectively.
In the \VBF- and \VH-tagged categories, both discriminants incorporate information from the \PH decay and
the two associated jets, and are denoted $\mathcal{D}_{\text{sig}}^{\text{EW}}$ and
$\mathcal{D}_{\text{int}}^{\text{EW}}$, respectively.

Alongside the primary observable $m_{4\Pell}$, using the two discriminants $\mathcal{D}_{\text{sig}}$ and
$\mathcal{D}_{\text{int}}$ would require a three-dimensional parameterization in the likelihood fit.
Such a high-dimensional space risks sparse population of many template bins in this analysis,
degrading the stability and reliability of the fit.
In earlier analyses this was addressed by merging sparsely populated template bins with their nearest
neighbors after template creation, a procedure used in Refs.~\cite{CMS:2024eka,CMS:2022ley}.
Nonetheless, such an approach remained ad hoc without a proper metric to optimize such a procedure.

The \textsc{MiLoMerge} procedure of Ref.~\cite{Davis:2026tpk} provides an optimal method to compress the
two-dimensional observable $(\mathcal{D}_{\text{sig}},\mathcal{D}_{\text{int}})$ into a single one-dimensional
discriminant, while preserving as much information as possible within a fixed number of bins.
The procedure iteratively merges bins globally down to a target, starting from an
initial fine binning, to minimize the loss from each merge; this results
in an optimal set of bins given a target number.
Because the event-level features that characterize off-shell production change with mass,
a separate set of 40 initial bins are defined for each \mell interval and optimize
the merging to maximize separation between signal, background, and interference.
With 21 \mell bins, the resulting binned representation achieves comparable
performance while using just over one-tenth the number of bins employed in Ref.~\cite{CMS:2024eka}.
Applying \textsc{MiLoMerge} thus preserves the benefits of a multidimensional approach while respecting the
statistical limits of the data.

\section{Implementation of fitting and associated uncertainties}\label{sec:SignalModelling}

We perform an extended maximum likelihood fit using the \textsc{Combine} tool~\cite{CMS:2024onh}
in which the probability density in the off-shell region is normalized to the total event yield in each category $k$
as a sum over all processes $j$ according to
\begin{widetext}    
\begin{equation}
    \label{eq:poffshell}
    \mathcal{P}_{jk}(\vec{x};\vec{\xi}_{jk},\vec\zeta) =
    {\mu_j^\text{off}}\mathcal{P}_{jk}^\text{sig} ( \vec{x};\vec{\xi}_{jk},\vec\zeta)
    + \sqrt{\mu_j^\text{off}}\mathcal{P}_{jk}^\mathrm{int} ( \vec{x};\vec{\xi}_{jk},\vec\zeta)
    + \mathcal{P}_{jk}^\text{bkg} ( \vec{x};\vec{\xi}_{jk})
    + \mu_j^\text{off} \frac{\Gref}{\GH}  \mathcal{P}_{jk}^\text{cross} (\vec{x};\vec{\xi}_{jk},\vec\zeta),
\end{equation}
\end{widetext}
where $\mu_j^{\text{off}}$ denotes the ratio of the observed off-shell cross section to the SM prediction,
$\vec{\xi}_{jk}$ are the nuisance parameters encoding the uncertainties in this parameterization,
$\vec{\zeta}$ represents the set of free parameters of interest that describe the kinematic distributions
for the process, and $\vec{x}$ represents the set of observables that enter the analysis.
The $\mathcal{P}_{jk}^{\text{sig}}$, $\mathcal{P}_{jk}^{\text{int}}$, $\mathcal{P}_{jk}^{\text{cross}}$, 
and $\mathcal{P}_{jk}^\text{bkg}$ probability densities
are normalized to the expected number of events, and are implemented as binned histograms (templates)
of the observables $\vec{x}$. The specific treatment of $\vec{\zeta}$ varies by analysis and is detailed in the dedicated
sections that follow, with the null hypothesis being the SM expectation.

The model explicitly includes contributions from the signal-background interference
(``int'') and a cross-feed component  (``cross''), which is described in the following paragraph. There are three signal production modes,
indexed by $j$: \ggH, EW production (including both \VBF and \VH), and $\PQq\PAQq$ production,
and three jet-tagged categories, indexed by $k$. 
All lepton flavors and data-taking periods are combined in the off-shell parameterization.
Pseudo-experiments were performed for the various fits shown below, and the nominal bounds of $-2\Delta\ln{\mathcal{L}}=0.99$
and $3.84$~\cite{Cowan:2010js,Feldman:1997qc} were found to sufficiently cover the 68 and 95\% confidence intervals.

In Eq.~(\ref{eq:poffshell}), \GH affects only the cross-feed arising from on-shell
Higgs boson production that is reconstructed in the off-shell region, where \Gref denotes the reference
\GH used in the simulation. Other processes can mimic off-shell Higgs boson production and decay
to four leptons in this mass range. In particular, we consider on-shell signal events in which the Higgs boson
decays to $2\Pell + \mathrm{X}$ alongside 2 associated leptons.
The dominant on-shell contamination of the off-shell region is the $\PZ(\Pell\Pell)\PH(2\Pell+\mathrm{X})$ process,
referred to as the \ZH cross-feed. This contribution is estimated using on-shell \ZH samples with
$\PH\to\PW\PW$ and $\PH\to\ZZ$ decays, allowing both hadronic and leptonic decays of the \PZ boson.
The main contributions come from the $\PH\to 2\Pell 2\PGn$ and $\PH\to 2\Pell 2\PQq$ final states produced
in association with $\PZ\to 2\Pell$.
Dedicated on-shell samples passed through full simulation are utilized for the cross-feed,
while cross-feed events from the combined on- and off-shell samples are removed
to avoid double counting events.
A smaller contribution from $\PQt\PAQt\PH$ production
is treated in the same manner.

The dominant background is the process $\PQq\PAQq\to\ZZ/\PZ\PGgst/\PGgst\PGgst\to 4\Pell$.
The $\Pg\Pg\to\ZZ/\PZ\PGgst/\PGgst\PGgst$ background and the EW backgrounds,
including vector boson scattering and $\PV\PV\PV$ processes, interfere with off-shell Higgs boson
production. An additional background, denoted \ZX below, arises when heavy-flavor hadron decays, 
nonprompt decays of light mesons inside jets, or charged hadrons
overlapping with \PGpz decays are misidentified as prompt leptons.
It is the only background without a corresponding signal process in Eq.~(\ref{eq:poffshell}),
and, therefore, there is no interference.

The \ZX background is estimated from data rather than simulation. Its dominant contribution
is $\PZ$+jets production, evaluated using control regions in data. These control regions require a lepton
pair satisfying all ${\PZ}_{1}$ candidate criteria plus two additional opposite-sign leptons that satisfy
looser identification requirements than those applied in the signal selection. The four leptons are then
required to pass the ${\PZ}_{1},{\PZ}_{2}$ selection. The background yield in the signal region is obtained
by weighting control-region events by the lepton misidentification probability $f_{\Pell}$, defined as the
probability that a nonprompt lepton, or a lepton not associated with the primary vertex, passes the
analysis selection. Further details of this method are given in Ref.~\cite{Sirunyan:2017exp}.

\begin{figure*}[!htbp]
    \centering
    \includegraphics[width=0.83\textwidth]{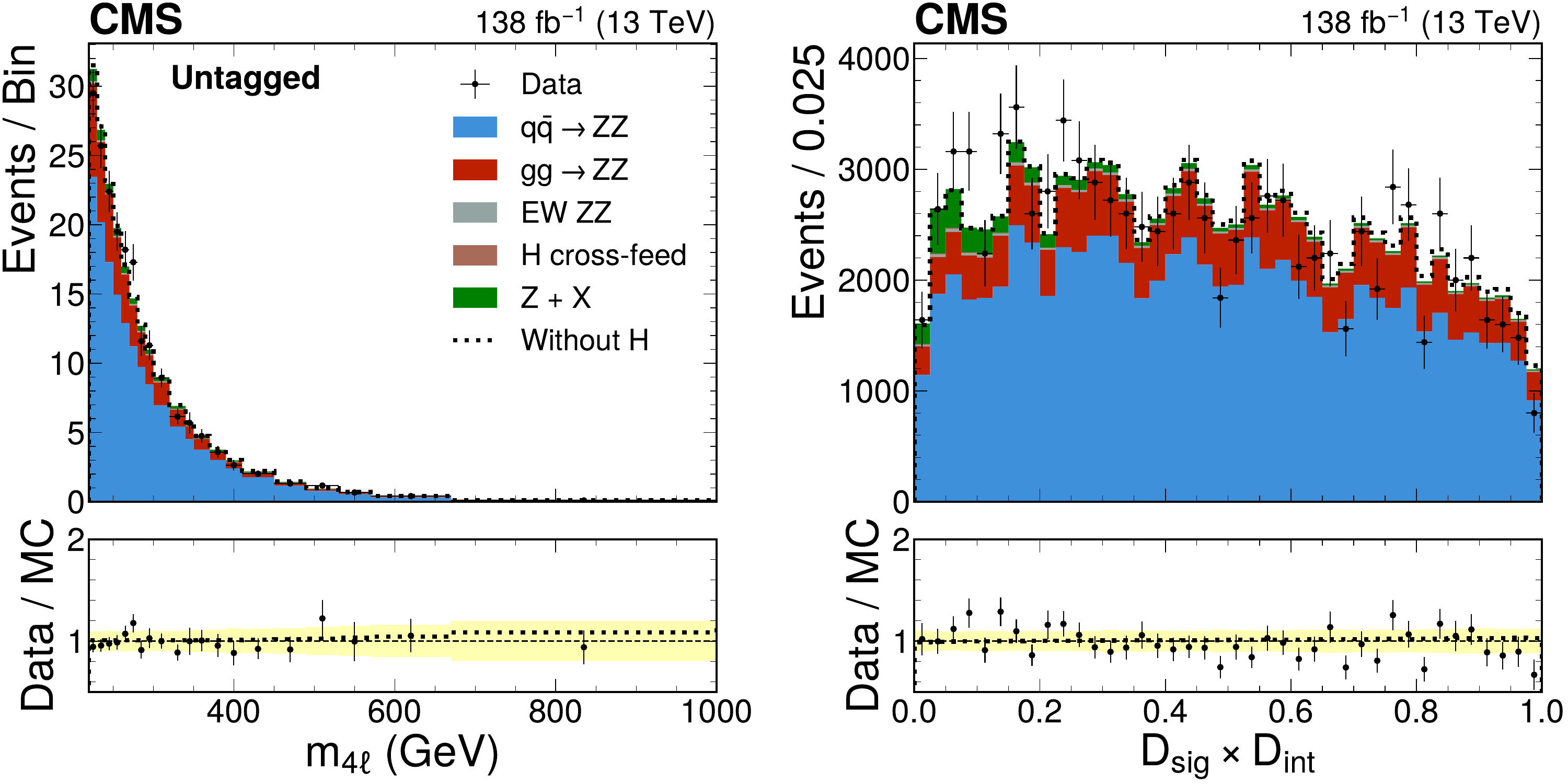}\\
    \includegraphics[width=0.83\textwidth]{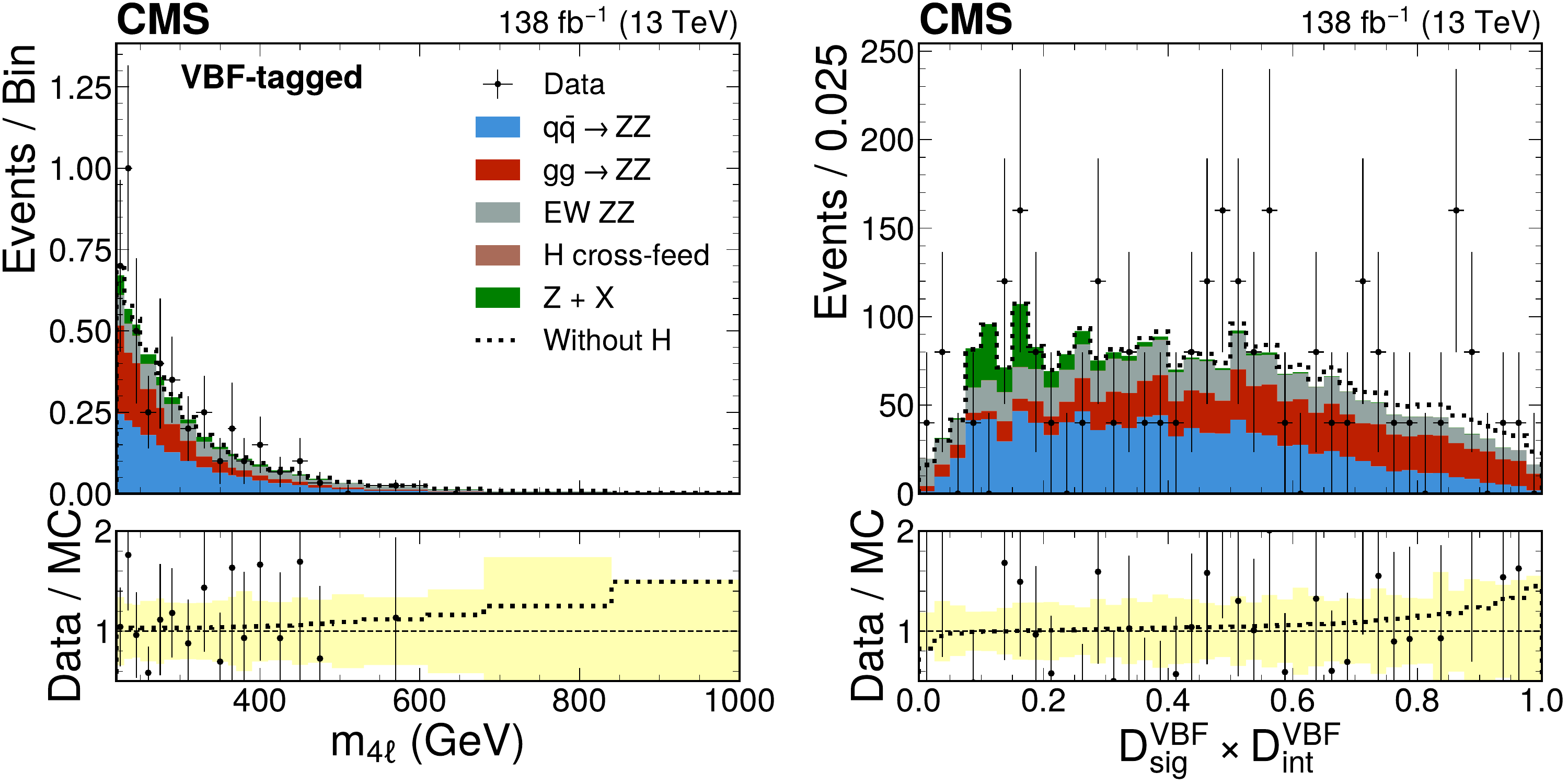}\\
    \includegraphics[width=0.83\textwidth]{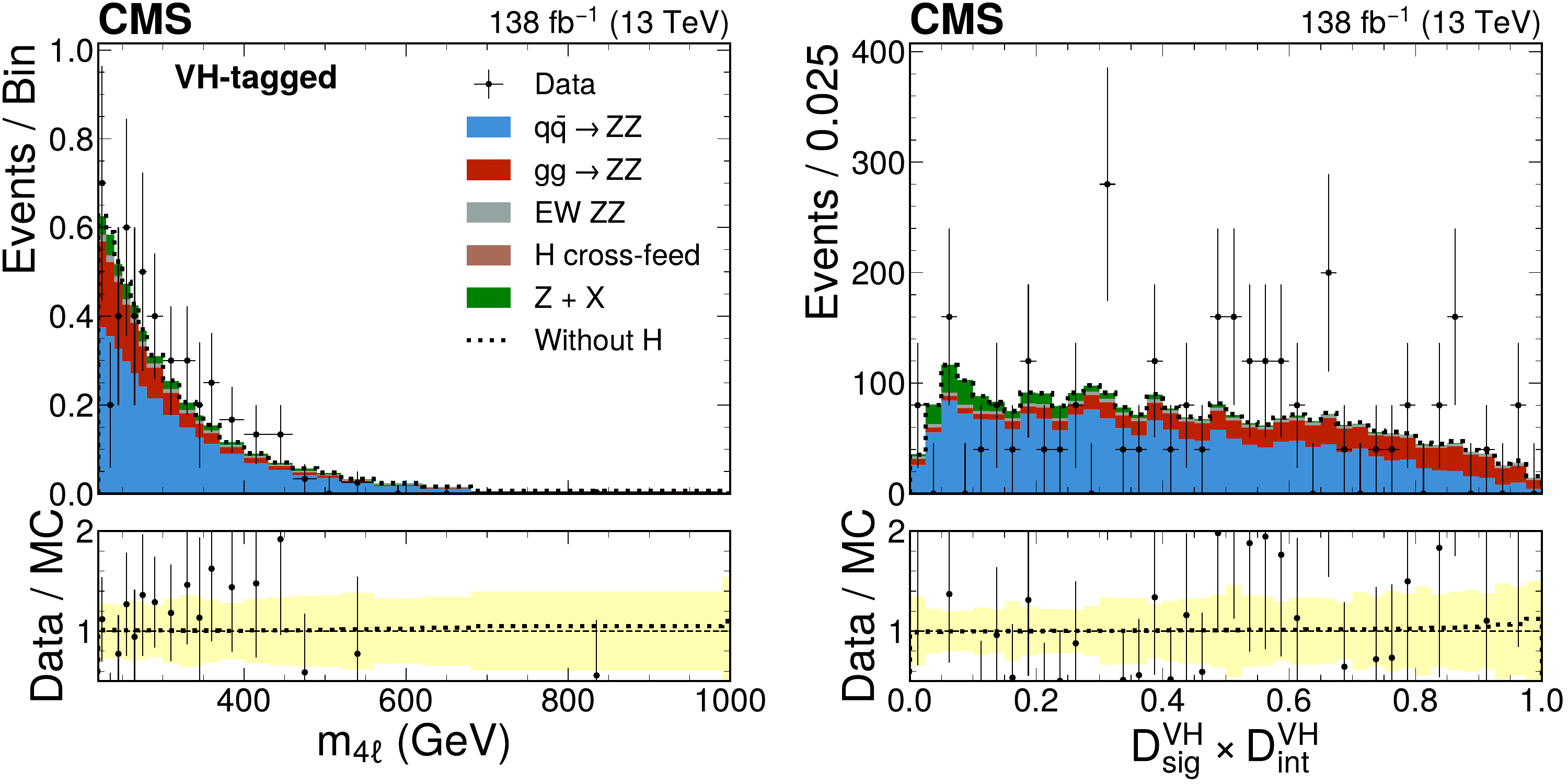}\\
    \caption{
        The pre-fit $m_{4\ell}$ distributions (left) and the combined discriminants (right) 
        in the off-shell region are shown for the
        untagged (upper), \VBF-tagged (middle), and \VH-tagged (lower) categories,
        divided by the bin width. The legend shows the expected
        signal, background, or their combined yield with interference for the various processes. The ratio of observation
        to expectation is displayed in the subpanel below the main plot with a ${\pm}1$ standard deviation uncertainty band in yellow.
        The dashed curve labeled ``Without $\PH$'' denotes the no-off-shell-$\PH$-production hypothesis.
    }\label{fig:SM_Expectation_plot}
\end{figure*}

In estimating the \ZX background, the flavor composition of jets misidentified as leptons can differ
between the $\PZ+1\Pell$ and $\PZ+2\Pell$ control regions. Combined with the statistical uncertainty in the $\PZ+2\Pell$ region, 
this results in an overall variation of about 30\% in the background yield.
The uncertainty in the misidentification rate modeling as a function of \pt and $\eta$,
together with the statistical uncertainty of the $\PZ+1\Pell$ control region, results in background-yield
uncertainties ranging from 32\% for the $4\Pe$ final state to 39\% for the $4\PGm$ final state.
Figure~\ref{fig:SM_Expectation_plot} shows the distributions of the two observable projections for the three
tagging categories, with the arbitrary bins of the discriminant sorted by the ratio of the no off-shell scenario to the SM.

The dominant systematic uncertainty in the off-shell measurements, represented by the nuisance parameters
$\vec{\xi}_{jk}$ in Eq.~(\ref{eq:poffshell}), arises from the modeling of the principal background,
$\PQq\PAQq\to\ZZ/\PZ\PGgst/\PGgst\PGgst\to 4\Pell$. Experimental uncertainties specific to the off-shell
analysis include those associated with jet energy calibration, which affect the event yields in the tagged
categories and have a larger impact in the \VBF- and \VH-tagged categories where the jets enter the
observables defined in Section~\ref{sec:Discriminants}.

Theoretical uncertainties affecting both the signal and background estimates include the effect of
missing higher orders in perturbative QCD and the choice of PDFs. The former is evaluated
by varying each of the
renormalization and factorization scales up and down by factors of 2 and 0.5 from their nominal value,
excluding cases in which the scales are varied in opposite directions. The uncertainty from the PDF set is taken
as the root-mean-square of the variations across the replicas of the default NNPDF ensemble
taken from LHAPDF~\cite{Whalley:2005nh,Bourilkov:2006cj}.
An additional $10\%$ uncertainty is assigned to the K-factor used for the $\Pg\Pg\to\ZZ$
prediction, reflecting the approximation that the K-factor for $\Pg\Pg\to\PH$ production
is also applicable to $\Pg\Pg\to\ZZ$ production~\cite{Caola:2016trd}.

The integrated luminosities for 2016, 2017, and 2018 carry individual systematic uncertainties
of 0.84--1.2\%~\cite{CMS:2021xjt,CMS-PAS-LUM-20-001},
while the combined uncertainty for the 2016--2018 data set is 0.73\%.
The uncertainty in the lepton identification, reconstruction, and selection efficiency ranges
from 2 to 14\% in terms of the overall event yield for the $4\PGm$ and $4\Pe$ final states,
respectively, and affects both signal and background processes.

\section{Probes of Higgs boson substructure}\label{sec:H_struct}

\begin{figure}[!htbp]
    \centering
    \includegraphics[width=0.49\textwidth]{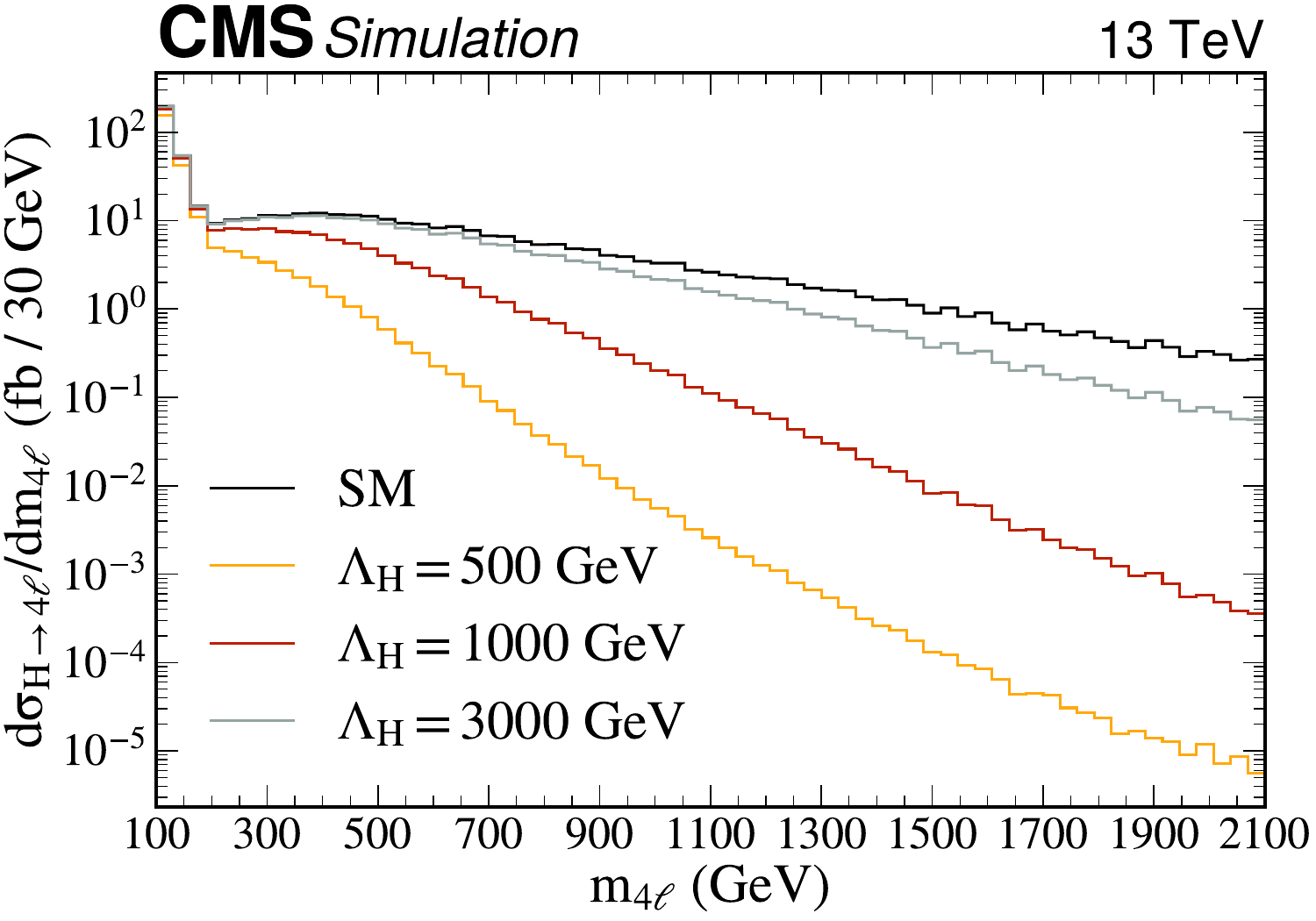}
    \includegraphics[width=0.49\textwidth]{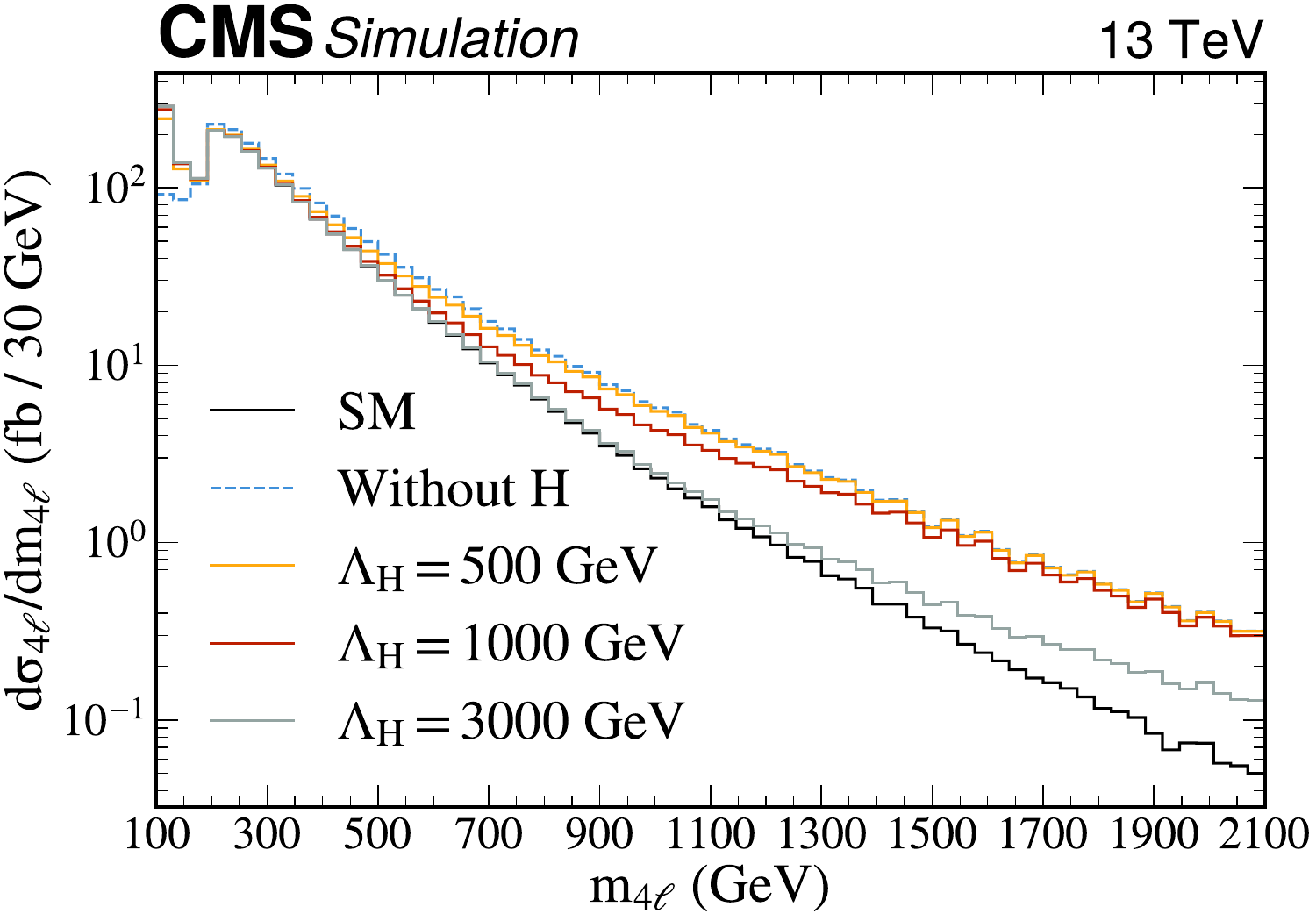}
    \caption{
Evolution of EW production of the Higgs boson as a function of \mell for selected values
of the compositeness scale $\Lambda_{\PH}$: $\PH$-only production (\cmsLeft) and the combined four-lepton
distribution including Higgs boson signal, background, and interference (\cmsRight).
The background-only scenario (``Without H'') is indicated by the dashed line on the right.
    }\label{fig:Struct_evolve}
\end{figure}

As noted in Section~\ref{sec:Introduction}, the composite Higgs boson scenario offers a potential
solution to the hierarchy problem. In this framework the Higgs boson is not elementary but is composed
of more fundamental constituents~\cite{Witzel:2019jbe}.
Equation~(\ref{eq:formfact-composite}) gives the form-factor that modifies the Higgs boson distribution
as a function of its four-momentum squared, $q^2$.
In the SM, where the Higgs boson is elementary, the associated compositeness scale $\Lambda_\PH$
is effectively infinite. Figure~\ref{fig:Struct_evolve} illustrates the impact of selected finite values
of $\Lambda_\PH$ on the \mell distribution for the EW process at the simulation level, shown both
for the inclusive four-lepton spectrum and for the Higgs boson signal.
The SM corresponds to $\Lambda_\PH=\infty$.

Note that in VBF production the Higgs boson can be produced through both $s$-channel and $t$-channel exchange 
(see Fig.~\ref{fig:prod_modes}). If only the $s$-channel contributed, the $q^2$ of the Higgs boson would be 
equal to $m_{4\ell}^2$, and a continuous parameterization in $\Lambda_\PH$ could be constructed. 
However, the presence of the $t$-channel breaks this one-to-one relation. 
At a given $m_{4\ell}^2$, the $t$-channel is less affected by the form-factor, since the Higgs boson can carry a smaller $q^2$.
Therefore, $q^2$ cannot be set equal to $\mell^2$, and it is not possible to determine $q^2$ event by event 
in the VBF topology. For this reason, $\Lambda_\PH$ is sampled at a discrete set of values.
Fourteen values of $\Lambda_\PH$ between 150 and 3000\GeV were chosen and the corresponding 
templates were generated by reweighting the samples described in Section~\ref{sec:Data}. 
An extended maximum likelihood fit is then performed to the on- and off-shell events using 
the categories and discriminant distributions given in Section~\ref{sec:Discriminants}.

In the on-shell region, the Higgs boson yield scales as $1/\Gamma_{\PH}$, whereas the off-shell region
is insensitive to $\Gamma_{\PH}$. Therefore, because the on-shell analysis depends on the absolute rate,
$\Gamma_{\PH}$ must be held fixed when probing $\Lambda_{\PH}$. By contrast, the off-shell approach
to probing Higgs boson substructure is relatively more model-independent: it does not require fixing $\Gamma_{\PH}$
and is implemented via a set of templates rather than through an overall rate. Finally, because of the complications
in determining the Higgs boson $q^2$ in off-shell VBF production discussed above,
the combined on- and off-shell analysis also proceeds at fixed values of $\Lambda_{\PH}$.

\begin{figure}[!htbp]
    \centering
    \includegraphics[width=\cmsFigWidth]{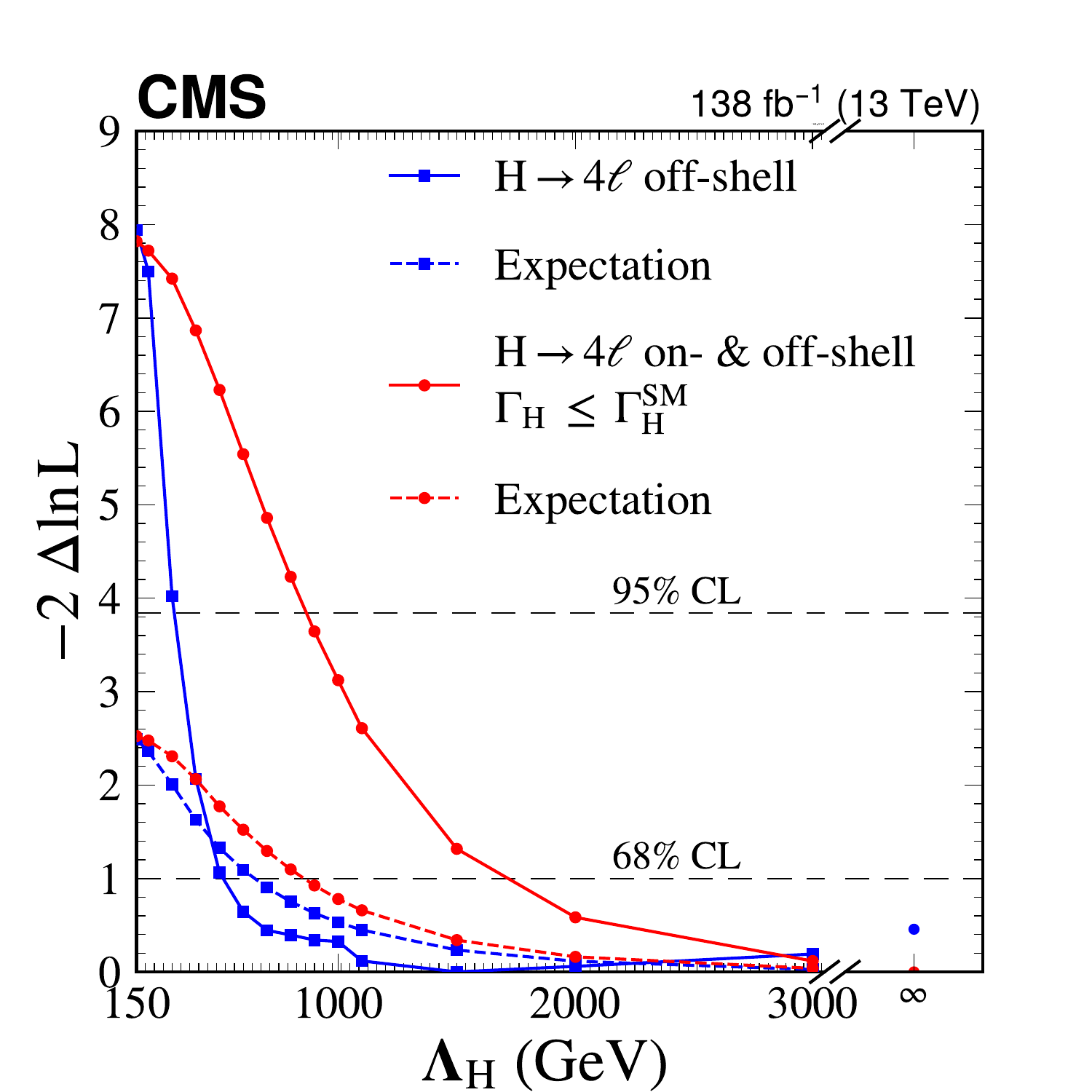}
    \caption{
       Observed (solid) and expected (dashed) profile-likelihood scans for $\Lambda_{\PH}$ from fits
       using off-shell production only (blue squares) and combined on- and off-shell production (red circles).
       Markers indicate the discrete $\Lambda_{\PH}$ points profiled.
       The black horizontal dashed lines mark the 68 and 95\% CL thresholds.
    }\label{fig:structScan}
\end{figure}

We consider two approaches to constrain possible values of $\Lambda_\PH$, both of which include an overall
signal strength scaling $\mu^\text{off}$ for Higgs boson production to allow for deviations in the data:
\begin{enumerate}
    \item
Use only off-shell Higgs boson production. This method probes $\Lambda_\PH$ without fixing $\Gamma_\PH$,
but it can yield unphysically large values of $\mu^\text{off}$ as $\Lambda_\PH$ decreases.
    \item
Use both on- and off-shell Higgs boson production. This reduces the tendency for $\mu^\text{off}$ to take unphysical
values, at the cost of fixing $\Gamma_\PH$ to the SM value of $\Gamma_{\PH}^\text{SM}=4.1\MeV$.
In fact, relaxing this constraint to $\Gamma_{\PH}\le \Gamma_{\PH}^\text{SM}$ yields the same results,
because the best fit point driven by the data will push $\Gamma_\PH$ to the largest allowed value
when $\mu^\text{off}$ is being profiled.
\end{enumerate}
The results are shown in Table~\ref{tab:structTab} and Fig.~\ref{fig:structScan}. The difference between
expectation and observation is consistent with a statistical fluctuation in the data, and the coverage of the 
confidence-level intervals is tested with simulated pseudo-experiments. This same fluctuation is
responsible for the difference between expectation and observation
in the significance of off-shell production discussed later in Section~\ref{sec:width}.
We also translate the 68\% CL (confidence level)
value of $\Lambda_{\PH}$ into a length scale in zeptometers ($1\unit{zm}= 10^{-21}\unit{m}$)
using Eq.~(\ref{eq:size-composite}).
Using only off-shell production, $\Lambda_{\PH}$ is constrained to be greater than 310\GeV at 95\% CL,
with the likelihood preferring a finite value near 1500\GeV.
When on- and off-shell data are combined while holding $\Gamma_{\PH}$
constrained below the SM value, the 95\% CL lower bound improves to 870\GeV.

\begin{table*}[!htb]
    \centering
    \topcaption{
    Summary of the observed and expected constraints on the parameters describing Higgs boson substructure, $\Lambda_{\PH}$ and $d$,
    obtained from $\PH\to\PZ\PZ\to4\Pell$. For each parameter the table lists the central value together with the 68 and 95\% CL intervals.
    The hyphens at the 95\% CL indicate that none of the tested hypotheses can be excluded.
    }
    \renewcommand{\arraystretch}{1.25}
\begin{scotch}{llllllll}
Parameter & Scenario & \multicolumn{3}{c}{Observed} & \multicolumn{3}{c}{Expected} \\
 & & Best fit & \multicolumn{1}{c}{68\% CL} & \multicolumn{1}{c}{95\% CL} & Best fit & \multicolumn{1}{c}{68\% CL} & \multicolumn{1}{c}{95\% CL} \\
\hline
$\Lambda_{\PH}$ (\GeVns) & off-shell & $1500$ & ${>}520$ & ${>}310$ & $\infty$ & ${>}650$ & \NA  \\
$\Lambda_{\PH}$ (\GeVns) & on- \& off-shell & $\infty$ & ${>}1720$ & ${>}870$ & $\infty$ & ${>}860$ & \NA   \\
$d$ (zm) & off-shell & $130$ & ${<}380$ & ${<}640$ & $0$ & ${<}300$ & \NA   \\
$d$ (zm) & on+off-shell & $0$ & ${<}120$ & ${<}230$ & $0$ & ${<}230$ & \NA   \\
\end{scotch}
     \label{tab:structTab}
\end{table*}

\section{Probes of heavy particles in Higgs boson production loops}\label{sec:H_heavy}
In this section we focus on Higgs boson couplings to heavy states, either the top quark or a hypothetical heavy quark $Q$,
which serves as a generic proxy for any new heavy particle entering the loop. We allow both \CP-even and -odd
couplings, as introduced in Eq.~(\ref{eq:LagrHqq}). Section~\ref{sec:H_Yukawa} treats Higgs boson Yukawa couplings to
light quarks. Both analyses share a common formalism and can therefore be regarded together as studies of the
Higgs Yukawa interactions with heavy and light fermions, as discussed below.

The Yukawa couplings of the Higgs boson to quarks govern several processes.
In off-shell $\PH$ production they enter primarily through the \ggH loop, which is mediated by quarks,
with a much smaller contribution from direct quark-antiquark annihilation.
In the SM this annihilation channel is negligible either because the relevant Yukawa coupling is very small
(for light quarks) or because their parton density in the proton is close to or at zero (for heavy states).
In \ggH, the top quark provides the dominant mediation, although theories predicting
rare contributions from either heavier or lighter states also exist.

In the off-shell region the probability parameterization follows
\begin{widetext}
\begin{align}
    \mathcal{P}_{k}^\text{off} = &\mathcal{P}_{k}^\text{bkg}
+ \frac{\Gamma_0}{\Gamma_\PH}\kappa_\PZ^2\left(
            \kappa_\PQt^2 \mathcal{P}_{\PQt\PAQt\PH,k}^\text{cross} +
            \kappa_\PZ^2 \mathcal{P}_{\PZ\PH\PZ\PZ,k}^\text{cross} +
            \kappa_\PW^2 \mathcal{P}_{\PZ\PH\PW\PW,k}^\text{cross}
            \right) \nonumber \\
     & + \kappa_\PW \kappa_\PZ^3 \mathcal{P}_{\text{EW},k}^{\text{sig},\PZ\PW}
    + \sum_{\PV=\PZ,\PW} \left[ \kappa_\PV^2 \kappa_\PZ^2 \mathcal{P}_{\text{EW},k}^\text{sig}
    + \kappa_\PV \kappa_\PZ \mathcal{P}_{\text{EW},k}^\text{int}\right]
    \nonumber \\
        &+ \sum_{\PQq=\PQu,\PQd,\PQs,\PQc} \left[
            \kappa_\PQq^2  \kappa_\PZ^2
            \left(
                \mathcal{P}_{\Pg\Pg,k}^{\text{sig},\PQq} + \mathcal{P}_{\PQq\PAQq,k}^{\text{sig},\PQq}
            \right) + \kappa_\PQq \kappa_{\PZ} \mathcal{P}_{\Pg\Pg,k}^\text{int,q}
        \right] \nonumber
        + \kappa_\PQb^2 \kappa_\PZ^2 \mathcal{P}_{{\Pg\Pg},k}^{\text{sig},\PQb} \nonumber \\
        &+ \sum_{\PQq} \quad \left[\sum_{\PQq^{\prime}=\PQu,\PQd,\PQs,\PQc,\PQb,\PQt,\PQQ,~\PQq^{\prime} \neq \PQq}
            \kappa_\PZ^2 \kappa_\PQq \kappa_{\PQq^{\prime}} \mathcal{P}_{{\Pg\Pg},k}^{\text{sig},\PQq\PQq^{\prime}}
            \right]
        + \sum_{\PQq^{\prime}=\PQu,\PQd,\PQs,\PQc,\PQt,\PQQ}\left[
            \kappa_\PZ^2 \kappa_\PQb \kappa_{\PQq^{\prime}} \mathcal{P}_{{\Pg\Pg},k}^{\text{sig},\PQb\PQq^{\prime}}
        \right] \nonumber \\
        &+ \sum_{\PQQ^{\prime}=\PQt,\PQQ} \left[
             \kappa_\PZ^2 \left(
                \kappa_{\PQQ^{\prime}}^2 \mathcal{P}_{{\Pg\Pg},k}^{\text{sig},\PQQ^{\prime}} +
                \widetilde{\kappa}_{\PQQ^{\prime}}^2 \mathcal{P}_{{\Pg\Pg},k}^{\text{sig},\widetilde{\PQQ}^{\prime}}
            \right)
            + \kappa_{\PQQ^{\prime}} \kappa_\PZ \mathcal{P}_{{\Pg\Pg},k}^{\text{int},\PQQ^{\prime}}
        \right] \nonumber \\
        &+ \kappa_\PQQ \kappa_\PQt \kappa_\PZ^2 \mathcal{P}_{{\Pg\Pg},k}^{\text{sig},\PQQ\PQt}
        + \widetilde\kappa_\PQQ \widetilde\kappa_\PQt \kappa_\PZ^2 \mathcal{P}_{{\Pg\Pg},k}^{\text{sig},\widetilde{\PQQ}\tilde{\PQt}} ,
        \nonumber \\
    \label{eq:off-shellYukawa}
\end{align}
\end{widetext}
where $\mathcal{P}$ denotes a particular normalized probability distribution represented by its own template;
$k$ indicates the jet-tagged category;
$\Pq$ labels the light quarks $\{\PQu,\PQd,\PQc,\PQs\}$;
$\Pq^{\prime}$ labels all \CP-even quarks $\{\PQu,\PQd,\PQc,\PQs,\PQb,\PQt,\PQQ\}$;
$\PQQ^\prime$ denotes the heavy states $\{\PQt,\,Q\}$; and
$\PV$ stands for the EW bosons $\PZ$ and $\PW$.
When custodial symmetry is assumed, $\kappa_{\PZ}=\kappa_{\PW}$.

In Eq.~(\ref{eq:off-shellYukawa}), $\mathcal{P}_{k}^{\text{bkg}}$ collects all background contributions that
do not depend on the Higgs boson couplings. The next three terms are cross-feed contributions (superscripted ``cross''),
corresponding to on-shell Higgs boson events produced in the $\PQt\PAQt\PH$ and $\PZ\PH$ modes with $\PH\to\PZ\PZ$
and $\PH\to\PW\PW$ that are reconstructed in the off-shell region with an incorrect lepton assignment.
These contributions are described in Section~\ref{sec:SignalModelling}, as well as Ref.~\cite{CMS:2024eka}.
The second line corresponds to EW production mediated by the $\PH\PZ\PZ$ or $\PH\PW\PW$ couplings,
including their interference.

The remaining terms in Eq.~(\ref{eq:off-shellYukawa}) account for contributions from both light-quark Yukawa
couplings ($\PQq=\PQu,\PQd,\PQs,\PQc$), the bottom quark, and heavy state Yukawa couplings ($Q^\prime=\PQt,\,Q$),
including their interference.
The third line describes \ggH and light-quark $\PQq\PAQq$
production, together with the interference between the \ggH signal and background.
The fourth line gives the interference between heavy- and light-quark loop contributions,
and the last two lines represent the \ggH contribution from heavy states.
The \CP-odd couplings $\widetilde{\kappa}_{\PQQ^\prime}$ enter only as squared terms because
their interference with \CP-even terms vanishes for the observables we consider,
which are not sensitive to \CP violation.

Reference~\cite{CMS:2024eka} examined the Higgs boson coupling to potential heavy state contributions
in the $\Pg\Pg\to \PH$ loop, adopting the $\kappa_{\PQQ}$ definition from Ref.~\cite{Davis:2021tiv}.
In this work we perform targeted scans over a set of \CP-even quark couplings in the \ggH loop
and their \CP-odd analogues, defined as in Ref.~\cite{Davis:2021tiv}:
$\kappa_{\PQt}$, $\kappa_{\PQQ}$, $\widetilde{\kappa}_{\PQt}$, and $\widetilde{\kappa}_{\PQQ}$.
The off-shell parameterization used here follows Eq.~(\ref{eq:off-shellYukawa}), with the light-quark terms
fixed to their SM values. Sensitivity to \CP-odd effects in the off-shell region arises from the absence
of interference between those \CP-odd contributions and the background, which characterizes the
off-shell Higgs boson signature in the \mell distribution.
While the \CP-even contribution cancels between the signal and interference,
the \CP-odd contribution does not.

\begin{figure*}[!htbp]
    \centering
    \includegraphics[width=0.45\textwidth]{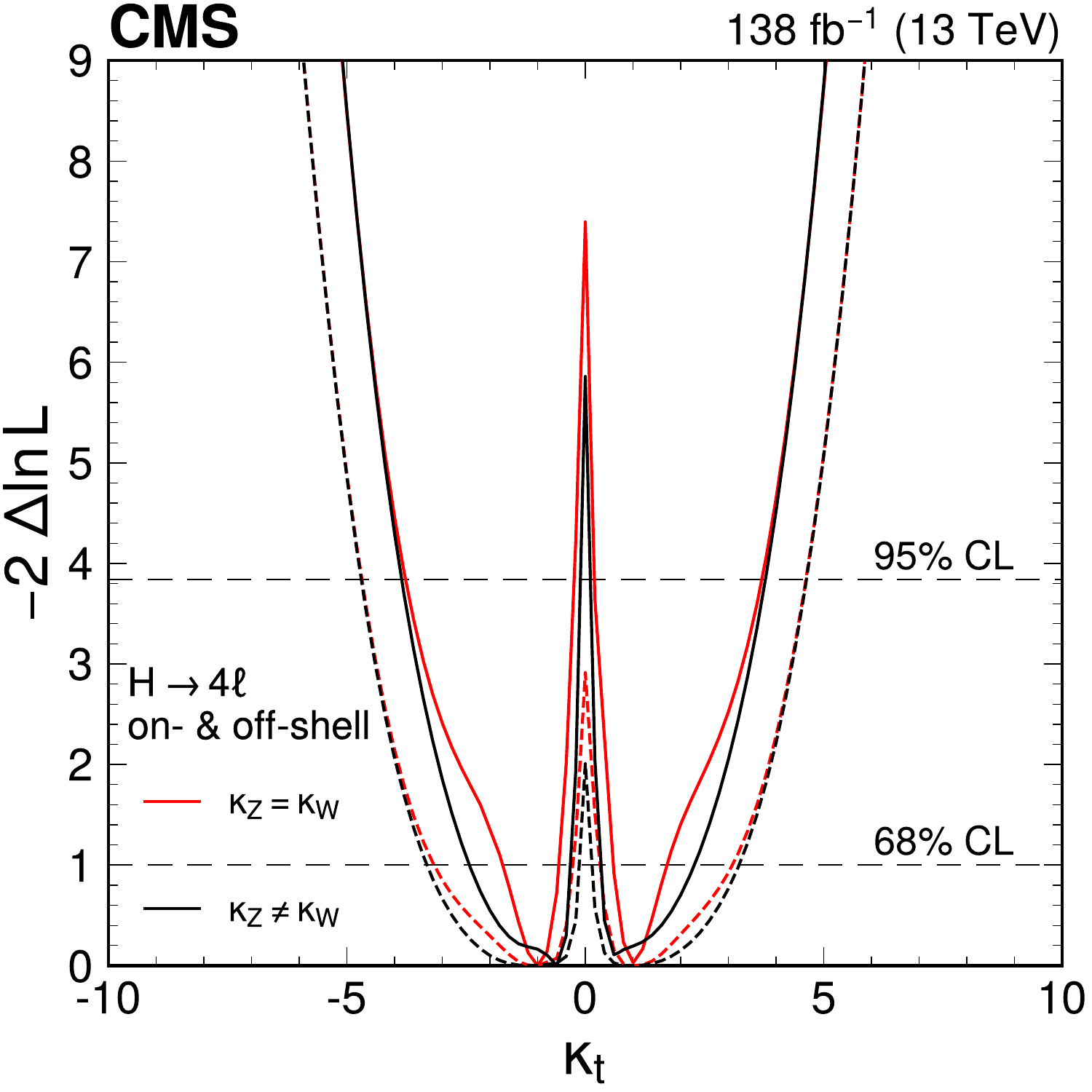}
    \includegraphics[width=0.45\textwidth]{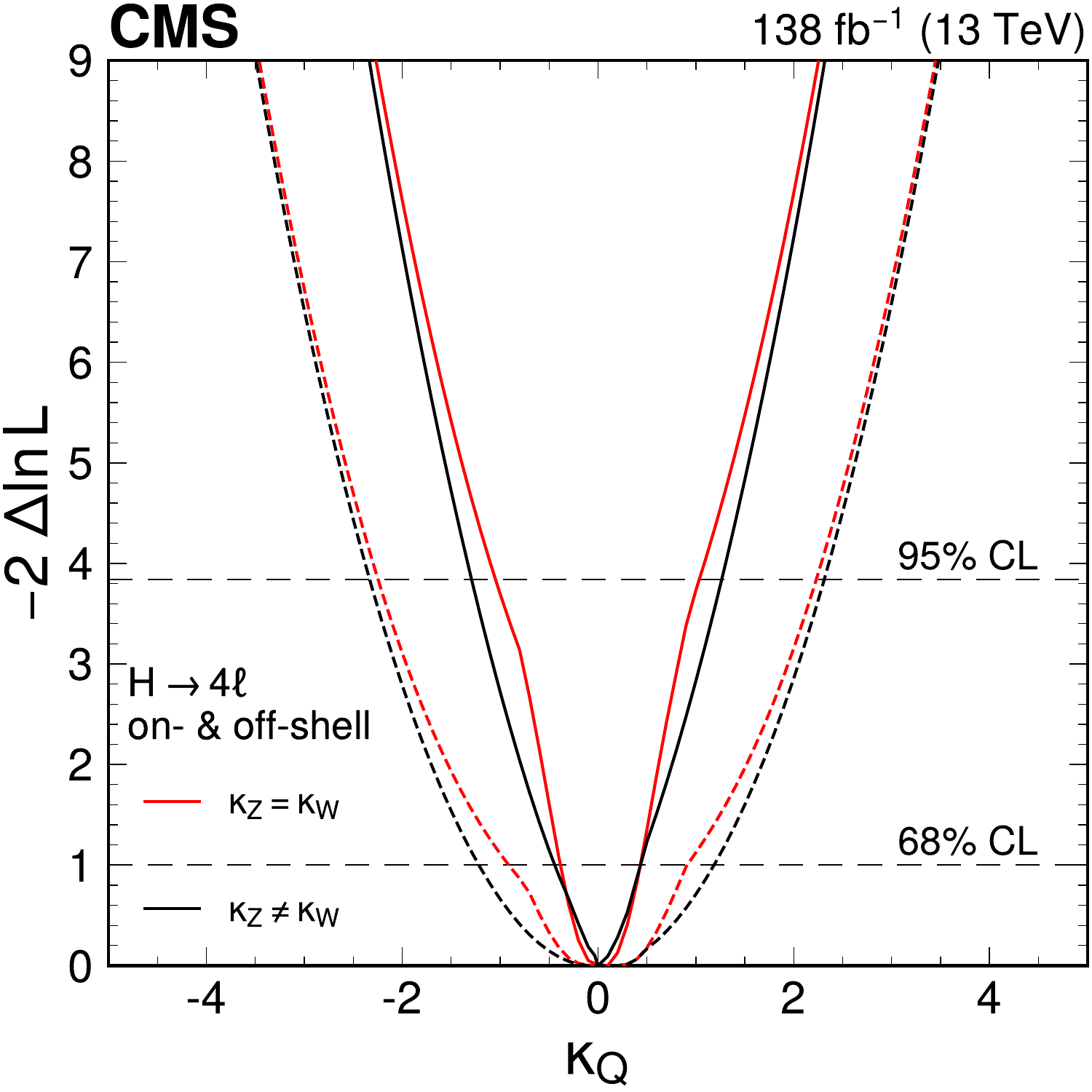}\\
    \includegraphics[width=0.45\textwidth]{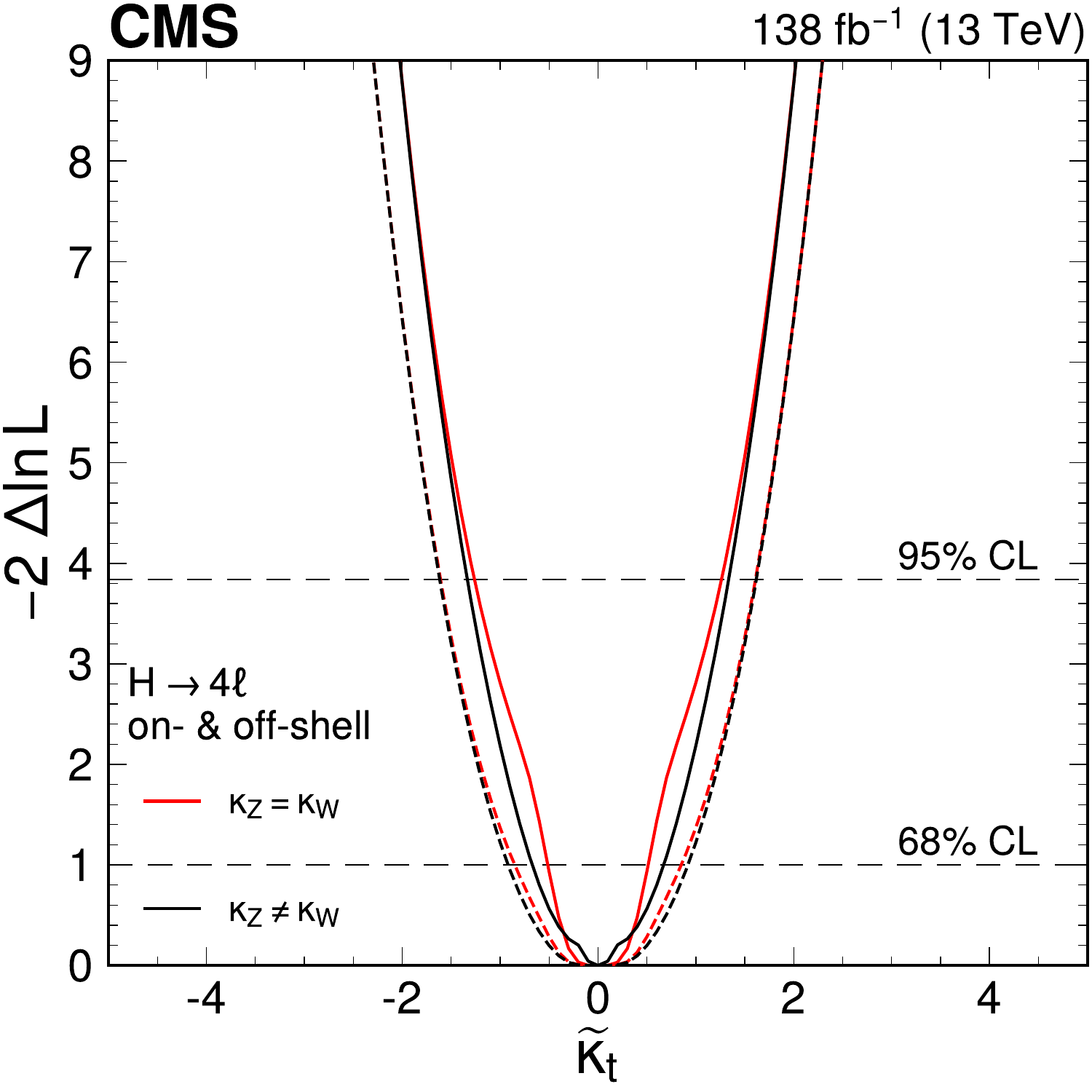}
    \includegraphics[width=0.45\textwidth]{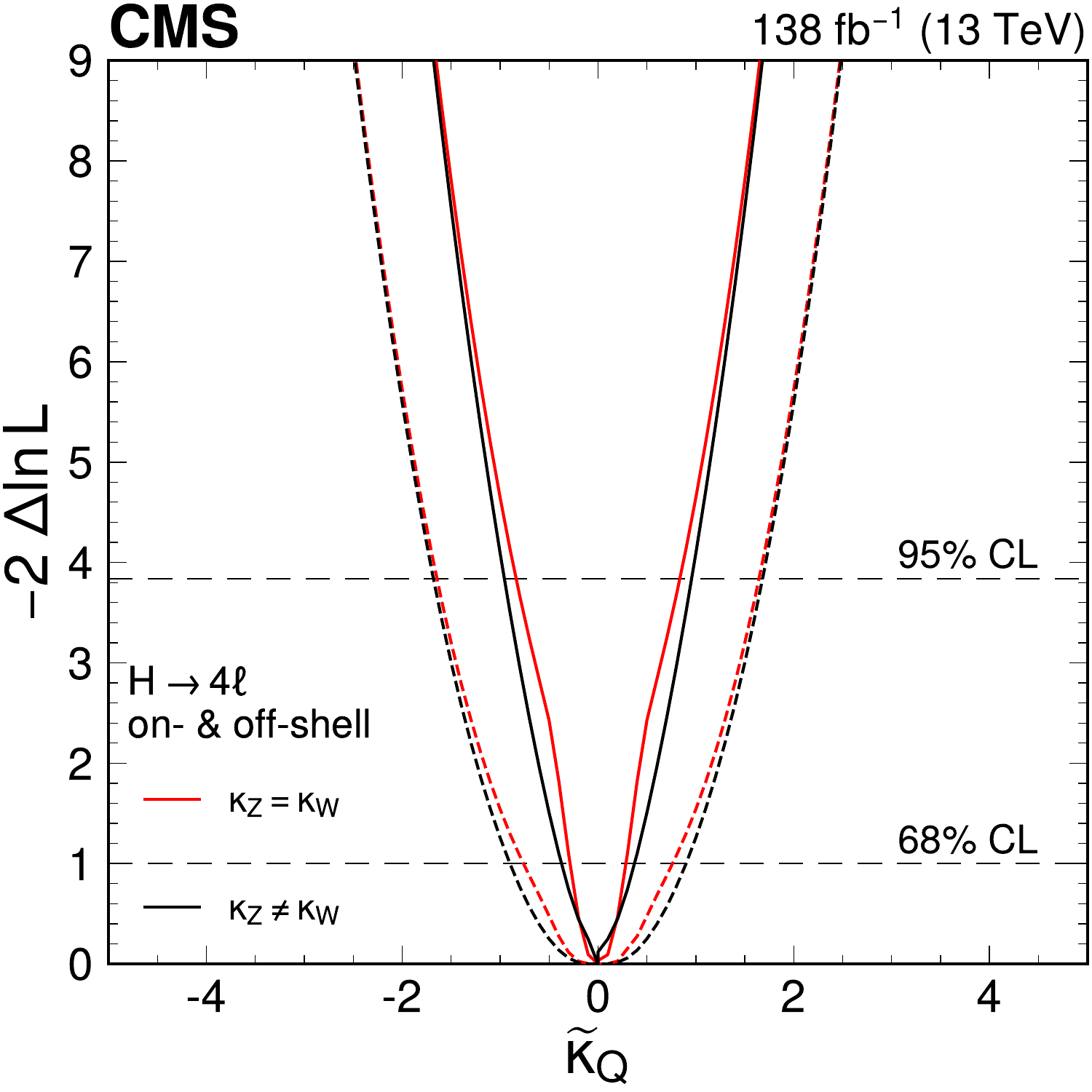}\\
\caption{
   Observed (solid) and expected (dashed) scans for the couplings $\kappa_{\PQt}$ (upper left),
   $\kappa_{\PQQ}$ (upper right), $\widetilde{\kappa}_{\PQt}$ (lower left), and $\widetilde{\kappa}_{\PQQ}$ (lower right),
   obtained from the combined on- and off-shell fit.
   All four couplings, together with $\kappa_{\PZ}$ and $\kappa_{\PW}$, are allowed to vary simultaneously.
   The horizontal dashed lines indicate the 68 and 95\% CL thresholds.
    }\label{fig:heavyYukawa_scan}
\end{figure*}

For consistency, the fit is performed simultaneously with the on-shell $\PH\to\PZ\PZ\to4\Pell$ analysis
of Ref.~\cite{CMS:2025xkn}, with $\kappa_{\PQb}$ and all light-quark Yukawa couplings fixed to their SM values.
However, $R_{\Pg\Pg}$, the parameterization of the \ggH cross section in the on-shell region, is expanded from
its construction in Ref.~\cite{CMS:2025xkn} to include effects from the heavy quarks measured in this section.
Scans are shown with and without the custodial symmetry assumption, the \GH is left free, 
and all relevant couplings are allowed to vary simultaneously.
Figure~\ref{fig:heavyYukawa_scan} shows the combined on- and off-shell profile scans for the
couplings $\kappa_{\PQt}$, $\kappa_{\PQQ}$, $\widetilde{\kappa}_{\PQt}$, and $\widetilde{\kappa}_{\PQQ}$,
with numerical results given in Table~\ref{table:heavyYukawa}.
The sign of $\kappa_{\PQt}$ is only weakly constrained by the interference in the \ggH loop,
as seen in the scan in Fig.~\ref{fig:heavyYukawa_scan}.
Resolving this sign ambiguity would require combining with analyses sensitive to the ttH vertex,
such as $\PQt\PH$ and $\PQt\PAQt\PH$ measurements.

The four $\kappa_{\PQt}$, $\kappa_{\PQQ}$, $\widetilde{\kappa}_{\PQt}$, and $\widetilde{\kappa}_{\PQQ}$
couplings map onto SM effective field theory (EFT) dimension-six operators that modify the top-quark Yukawa interactions
$(\mathcal{O}_{uH}$, $\mathcal{O}_{HD}$, $\mathcal{O}_{H\Box}$),  
and, through loop matching, to effective \PH-gluon contact operators
$\mathcal{O}_{HG}$ and $\widetilde{\mathcal{O}}_{HG}$~\cite{Dedes:2017zog,Davis:2021tiv}.
The relevant Wilson coefficient $C_{uH}$ is complex: its real part produces a \CP-even shift of the Yukawa
(mapped to $\kappa$) while its imaginary part induces a \CP-odd (pseudoscalar) component (mapped to $\widetilde{\kappa}$).
Thus, this analysis is performed in an EFT framework, with four real degrees of freedom modifying the \ggH loop.

\begin{table*}[!htbp]
    \centering
    \topcaption{
       Summary of the heavy-quark Yukawa coupling measurements for $\PH\to\PZ\PZ\to4\Pell$,
       showing central values with 68\% CL uncertainties and 95\% CL intervals (in square brackets),
       obtained from the combined on- and off-shell fit in which all heavy-quark couplings are allowed to vary
       simultaneously together with $\kappa_{\PZ}$ and $\kappa_{\PW}$. 
     }
    \renewcommand{\arraystretch}{1.25}
    \begin{scotch}{llllll}
        $\kappa_{\PQQ^\prime}$   &  Scenario &  \multicolumn{2}{c}{Observed} & \multicolumn{2}{c}{Expected} \\
        & & \multicolumn{1}{c}{68\% \CL} & \multicolumn{1}{c}{95\% \CL} & \multicolumn{1}{c}{68\% \CL} & \multicolumn{1}{c}{95\% \CL} \\
        \hline
        $\kappa_\PQt$   & $\kappa_{\PZ} = \kappa_{\PW}$          & $-1.00^{+0.44}_{-0.73}$ & $[-3.77,\,-0.14]$ & $1.00^{+2.08}_{-0.79}$ & $[-4.69,\, 4.63]$\\
        & & $\cup (0.60, 1.73)$ & $\cup [0.12,\,3.70]$ & $\cup (-3.08, -0.21)$ &   \\
        $\kappa_\PQt$   & $\kappa_{\PZ} \neq \kappa_{\PW}$          & $-0.60^{+0.28}_{-0.83}$ & $[-3.85,\,-0.10]$ & $1.00^{+2.22}_{-0.90}$ & $[-4.70,\, 4.64]$\\
        & & $\cup (0.28, 2.23)$ & $\cup [0.10,\,3.85]$ & $\cup (-3.32, -0.19)$ &   \\

        $\kappa_{\PQQ}$      & $\kappa_{\PZ} = \kappa_{\PW}$          & $0.10^{+0.33}_{-0.48}$ & $[-1.05,\,1.03]$ & $0.00 \pm 0.91$ & $[-2.35,\,2.23]$\\
        $\kappa_{\PQQ}$      & $\kappa_{\PZ} \neq \kappa_{\PW}$          & $0.00^{+0.43}_{-0.44}$ & $[-1.29,\,1.26]$ & $0.0 \pm 1.2$ & $[-2.3,\,2.3]$\\

        $\widetilde{\kappa}_\PQt$ & $\kappa_{\PZ} = \kappa_{\PW}$ & $0.00\pm0.51$ & $[-1.26,\,1.26]$ & $0.00\pm0.85$ & $[-1.61,\,1.61]$\\
        $\widetilde{\kappa}_\PQt$ & $\kappa_{\PZ} \neq \kappa_{\PW}$ & $0.00\pm0.67$ & $[-1.33,\,1.33]$ & $0.00\pm0.92$ & $[-1.62,\,1.62]$\\

        $\widetilde{\kappa}_\PQQ$    & $\kappa_{\PZ} = \kappa_{\PW}$ & $0.00\pm0.28$ & $[-0.84,\,0.84]$ & $0.00\pm0.76$ & $[-1.64,\,1.64]$\\
        $\widetilde{\kappa}_\PQQ$    & $\kappa_{\PZ} \neq \kappa_{\PW}$ & $0.00\pm0.37$ & $[-0.96,\,0.96]$ & $0.00\pm0.90$ & $[-1.68,\,1.68]$\\
    \end{scotch}\label{table:heavyYukawa}
\end{table*}

\section{Constraints on Higgs boson Yukawa couplings to light quarks}\label{sec:H_Yukawa}

\begin{figure}[!bh]
    \centering
    \includegraphics[width=0.35\textwidth]{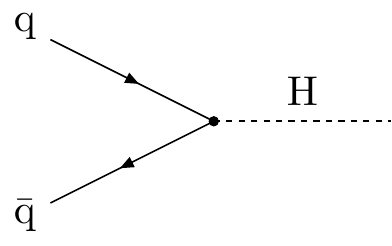}
    \caption{
    Leading-order Feynman diagram illustrating the direct quark-antiquark annihilation channel for Higgs boson production.
     }\label{fig:prod_qqbarH}
\end{figure}

In this section we concentrate on Higgs boson Yukawa couplings to light quarks ($\PQq=\PQu,\PQd,\PQs,\PQc$).
We probe the \CP-even term in Eq.~(\ref{eq:LagrHqq}) as a generic proxy for light-quark contributions to the
\ggH loop. At the same time, a production channel that is negligible in the SM becomes relevant here:
direct quark-antiquark annihilation, $\PQq\PAQq\to\PH$, illustrated in Fig.~\ref{fig:prod_qqbarH}.

The analysis follows the strategy recently used in the on-shell Higgs boson study~\cite{CMS:2025xkn},
which utilized the approach of Ref.~\cite{Zhou:2015wra}, but here we examine the full off-shell differential distribution
rather than only the on-shell cross section.
Results are presented for the combined analysis of both regions, using the $\PH\to\PZ\PZ\to4\ell$ channel.

The cross section of the on-shell Higgs boson production at the LHC, $\Pp\Pp\to\PH\to4\Pell$,
is inversely proportional to \GH and is parameterized as shown below~\cite{CMS:2025xkn}:
\begin{widetext}    
\begin{equation}
    \label{eq:ratio-sigma}
    \sigma_{\Hell}=
    \frac{\Gamma_{\PH\to4\ell}^\text{SM} \, \kappa_{\PZ}^2}{\Gamma_\PH(\kappa_{\PQu,\PQd,\PQs,\PQc})}
    \left(
   R_{\Pg\Pg}(\kappa_{\PQu,\PQd,\PQs,\PQc})  \, \sigma^\text{SM}_{\ggH}
    +\sum_{\PQq}\kappaq^2 \, \sigma^\text{SM}_{\PQq\PAQq\to \PH}
    +\sigma^\text{SM}_{\PQt\PAQt\PH}
    +\sigma^\text{SM}_{\PQt\PH}
    +\sum_{\PV}\kappa_{\PV}^2 \, \sigma^\text{SM}_{\PV\PV\PH}
    \right),
\end{equation}
where \GH is parameterized as
\begin{equation}
    \label{eq:ratio-width}
    \GH =
    R_{\Pg\Pg}(\kappa_{\PQu,\PQd,\PQs,\PQc})  \, \Gamma^\text{SM}_{\PH\to\Pg\Pg}
    +\sum_{\PQq=\PQu,\PQd,\PQs,\PQc,\PQb}\kappaq^2 \, \Gamma^\text{SM}_{\PH\to\PQq\PAQq}
    +\sum_{\PV\PV^\prime}\kappa_{\PV(\PV^\prime)}^2 \, \Gamma^\text{SM}_{\PH\to\PV\PV^\prime}
    +\sum_{\Pell}\Gamma^\text{SM}_{\PH\to\Pell\Pell}
    +\Gamma_\text{BSM}.
\end{equation}
\end{widetext}
The partial widths $\Gamma^{\mathrm{SM}}_{\PH\to f}$ are computed with SM values for all couplings,
while the partial width to BSM final states is unknown a priori and is only constrained by
$\Gamma_{\PH}^{\mathrm{BSM}}\ge 0$.
In the partial width for $\PH\to\PV\PV'$, $\PV$ and $\PV'\in\{\PW,\PZ,\PGg\}$.
Because these decays are dominated by the tree-level $\PH\PZ\PZ$ and $\PH\PW\PW$ couplings,
effects from light-quark Yukawa modifiers are highly suppressed and negligible compared with a direct
$\PH\to\PQq{\PAQq}$ partial width when light-quark couplings are enhanced.
Consequently, the dependence on $\kappa_{\PQq}$ appears only in the $\PH\to\PQq{\PAQq}$ and $\PH\to\Pg\Pg$ partial widths.
The \ggH production rate in Eq.~(\ref{eq:ratio-sigma}) and the partial width for $\PH\to\Pg\Pg$ in Eq.~(\ref{eq:ratio-width})
are both multiplied by the same factor $R_{\Pg\Pg}$, which encodes the dependence on the couplings of all quark types
entering the $\PH\to\Pg\Pg$ loop. Further details of calculations are given in Ref.~\cite{CMS:2025xkn}.

In the on-shell analysis~\cite{CMS:2025xkn} the constraint $\abs{\kappa_{\PZ}}\leq 1$ was imposed
and custodial symmetry was assumed, $\kappa_{\PZ}=\kappa_{\PW}$.
While this is well motivated in many theoretical scenarios~\cite{LHCHiggsCrossSectionWorkingGroup:2013rie},
it is a model-dependent assumption. Without it, a reduced on-shell cross section could be masked
by an arbitrarily large $\kappa_{\PZ}^2$ in Eq.~(\ref{eq:ratio-sigma}).
The off-shell parameterization, being independent of $\Gamma_{\PH}$ and exploiting shape information
in the differential distributions, allows us to drop that constraint. The off-shell measurement thus helps
lift degeneracies in $\kappa_{\PZ}$, while the on-shell data primarily determine the sensitivity to the
Higgs boson couplings to light quarks.

The off-shell region is parameterized by Eq.~(\ref{eq:off-shellYukawa}) with heavy state parameters and the
bottom quark fixed to their
SM values ($\kappa_{\PQb}=\kappa_{\PQt}=1$, $\widetilde{\kappa}_{\PQt}=\kappa_{\PQQ}=\widetilde{\kappa}_{\PQQ}=0$).
This parameterization resembles Eq.~(\ref{eq:ratio-sigma}) but uses differential templates of observables rather than
integrated cross sections. The \GH does not enter the off-shell treatment,
and the $\PQt\PAQt\PH$ and $\PQt\PH$ production modes are negligible in the off-shell region.
Templates are produced for each process, signal, background, and their interference, in each tagged category, as specified
 in Eq.~(\ref{eq:off-shellYukawa}).

\begin{figure*}[!htbp]
    \centering
    \includegraphics[width=0.49\textwidth]{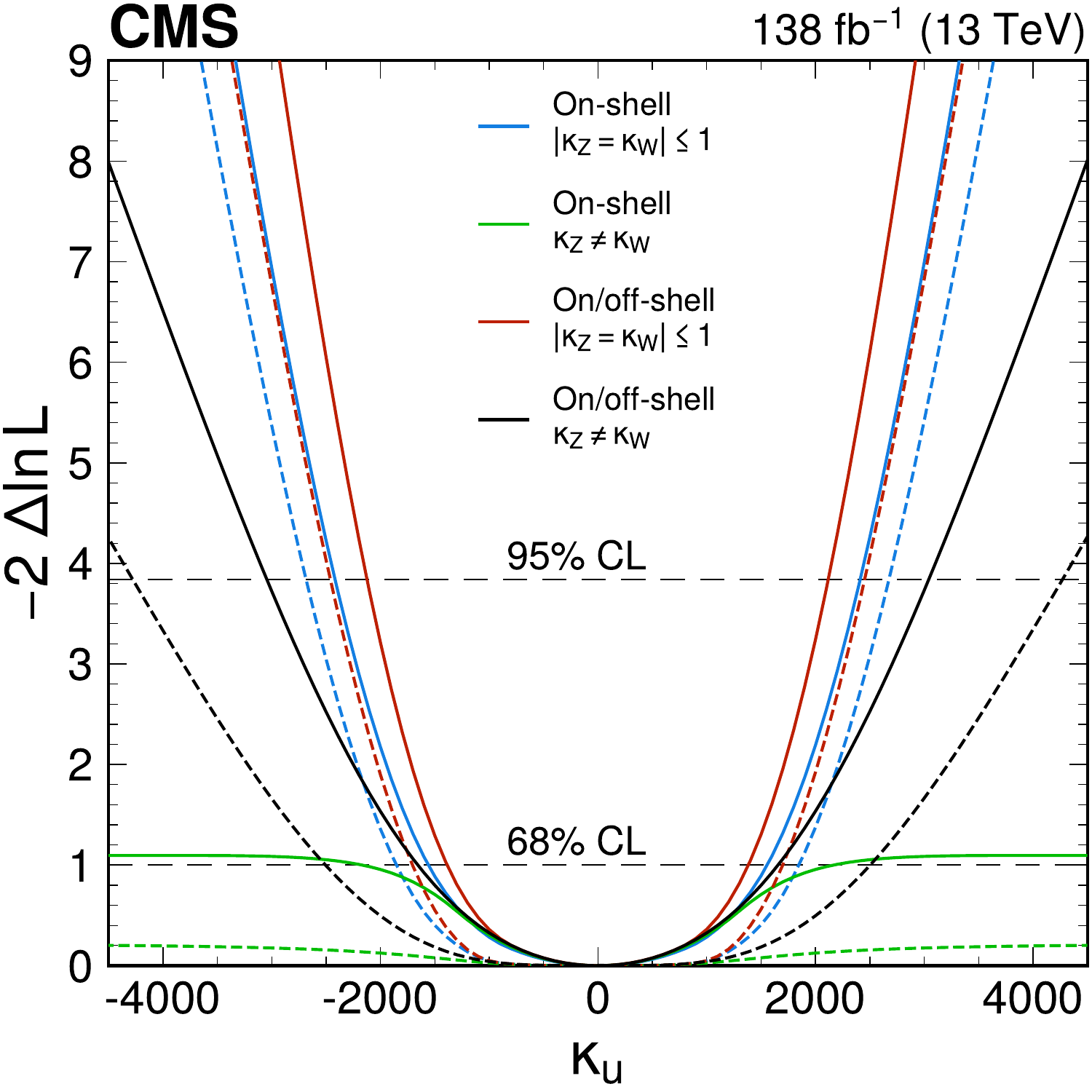}
    \includegraphics[width=0.49\textwidth]{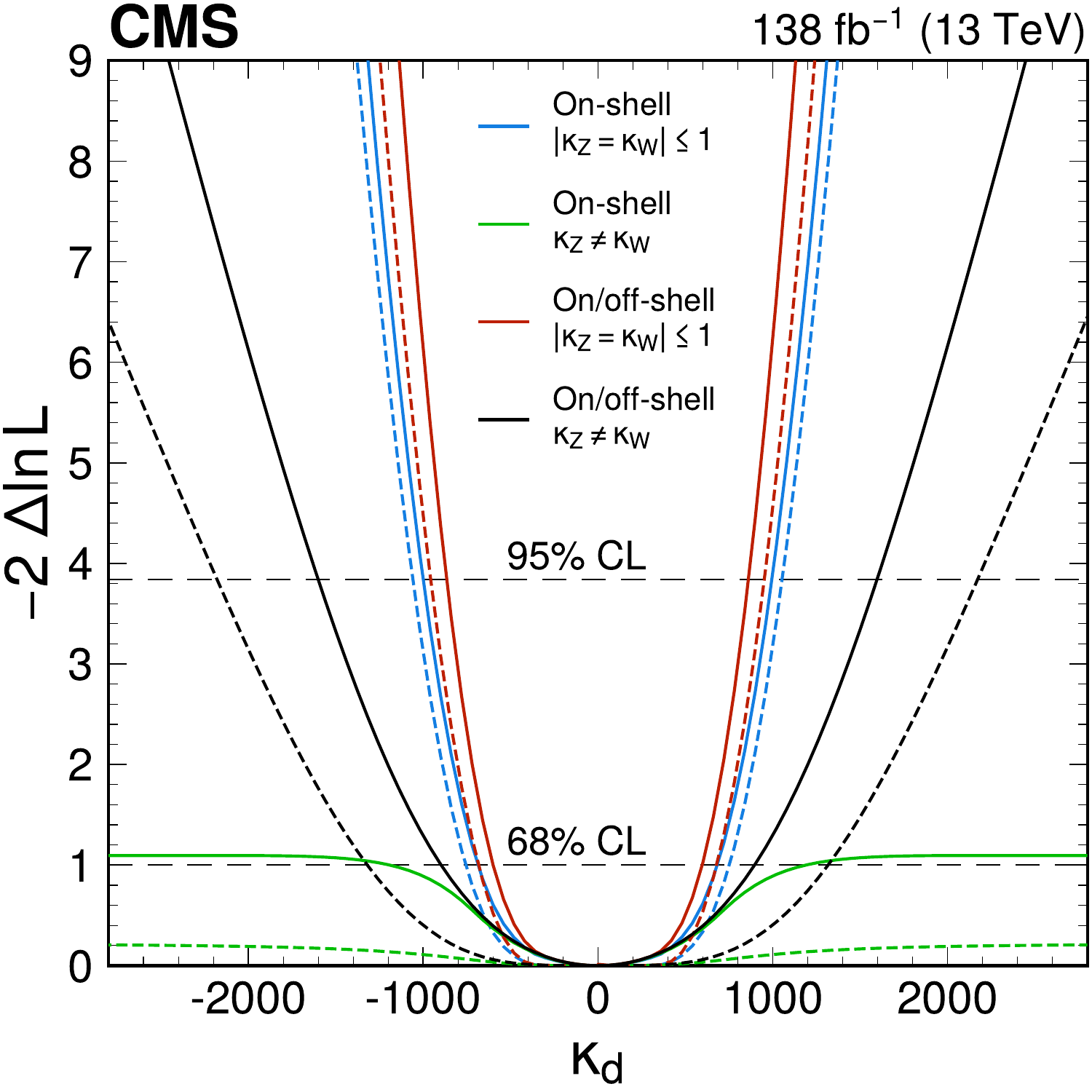}\\
    \includegraphics[width=0.49\textwidth]{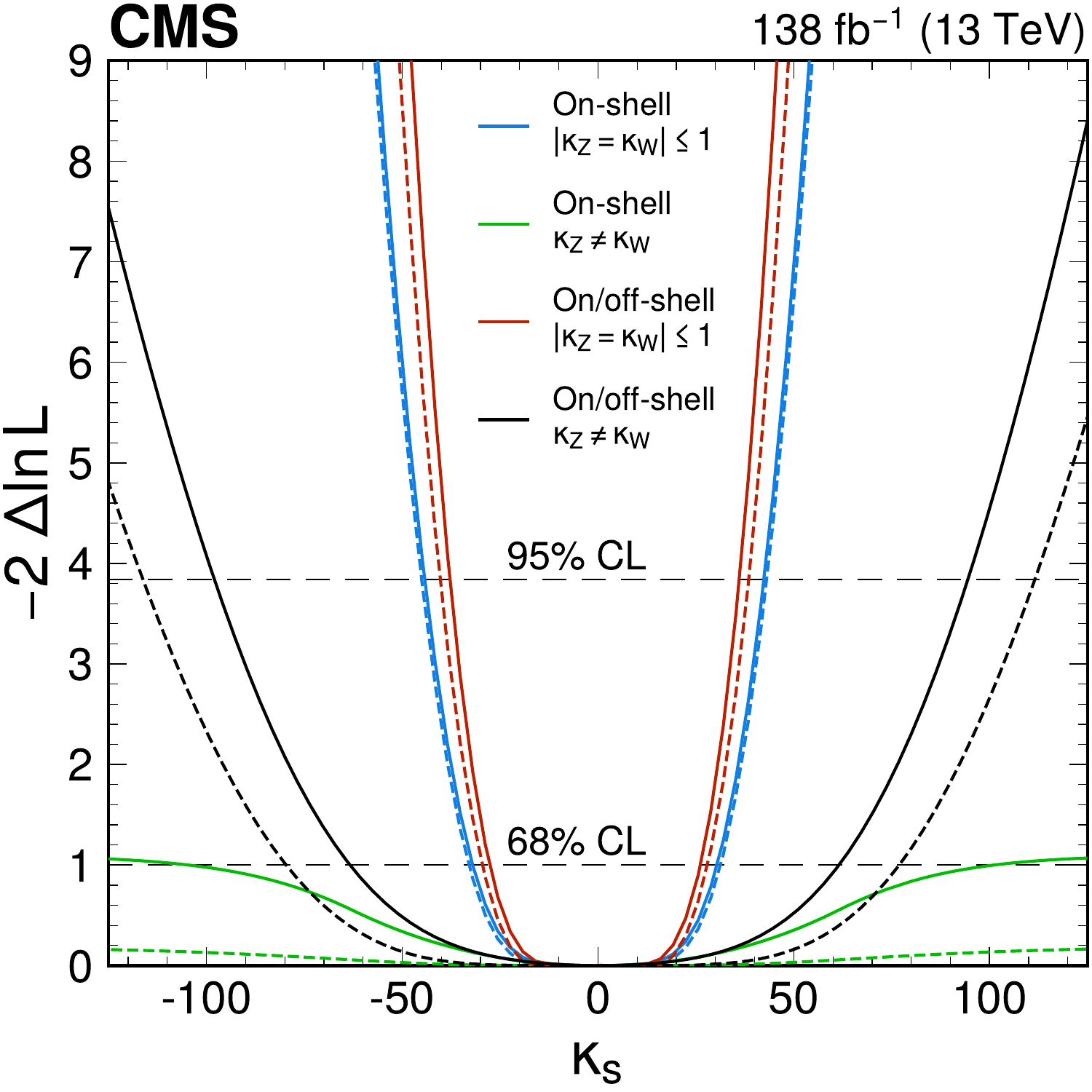}
    \includegraphics[width=0.49\textwidth]{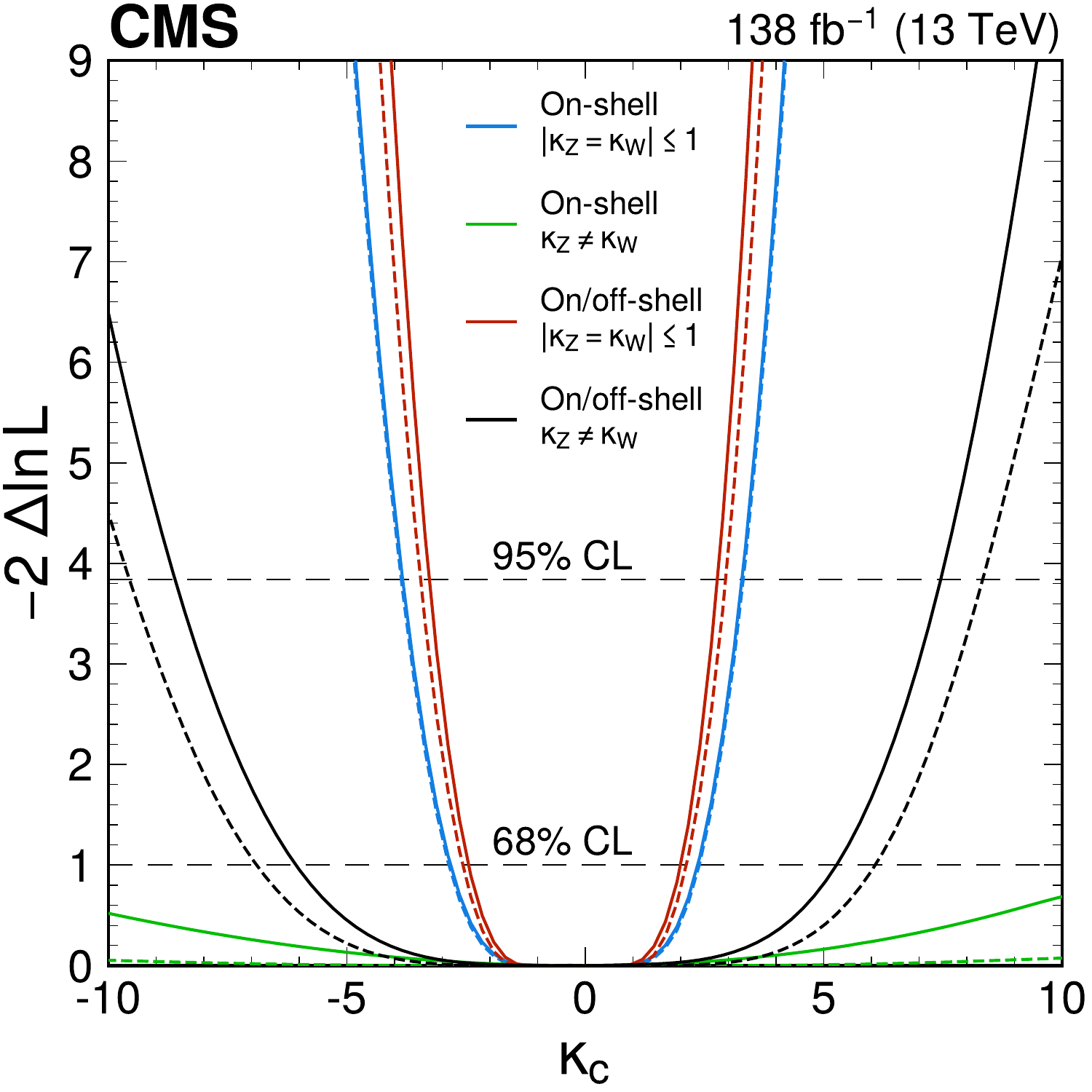}\\
\caption{
  Observed (solid) and expected (dashed) profile scans of the Higgs boson coupling modifiers to the light quarks:
  $\kappa_{\PQu}$ (up, upper left), $\kappa_{\PQd}$ (down, upper right), $\kappa_{\PQs}$ (strange, lower left),
  and $\kappa_{\PQc}$ (charm, lower right), shown for four analysis configurations: on-shell only or combined
  on- and off-shell production, and with or without the assumptions $\abs{\kappa_{\PZ}}\leq 1$ and
  $\kappa_{\PZ}=\kappa_{\PW}$. In each scan the other light-quark couplings are profiled simultaneously.
  The black horizontal dashed lines mark the 68 and 95\% \CL thresholds.
    }\label{fig:yukawa_scan}
\end{figure*}

Figure~\ref{fig:yukawa_scan} shows the combined on- and off-shell profile scans for
$\kappa_{\PQu,\PQd,\PQs,\PQc}$ Yukawa couplings.
Numerical results are given in Table~\ref{table:lightYukawa}.
The values quoted in Table~\ref{table:lightYukawa} are obtained without imposing custodial symmetry,
while scans performed under the original assumptions of Ref.~\cite{CMS:2025xkn} are also displayed
in Fig.~\ref{fig:yukawa_scan} for comparison.
The parameters $\barkappaq$ are presented in Table~\ref{table:lightYukawa_kappaBar}.

As shown in Fig.~\ref{fig:yukawa_scan}, removing the on-shell assumptions $\abs{\kappa_{\PZ}}\leq 1$ and
$\kappa_{\PZ}=\kappa_{\PW}$ removes any 95\% CL constraint (green curve). Including the off-shell information
(black curve) restores limits to a level comparable to the on-shell result obtained with those assumptions,
although the combined analysis yields only a modest improvement (red curve) relative to the on-shell fit
(blue curve) when those assumptions are imposed.
We therefore conclude that the principal benefit of including the off-shell analysis in the fit to the light-quark
Yukawa couplings is that it allows us to drop the on-shell assumptions $\abs{\kappa_{\PZ}}\leq 1$ and
$\kappa_{\PZ}=\kappa_{\PW}$ while retaining meaningful constraints.

In the SMEFT interpretation, the light-quark Yukawa modifiers map onto the dimension-six Yukawa operators
(e.g., $\mathcal{O}_{uH}$ for up-type and $\mathcal{O}_{dH}$ for down-type quarks) for the four flavors
considered ($\PQu,\PQd,\PQs,\PQc$)~\cite{Erdelyi2025}. Each operator's Wilson coefficient encodes a \CP-even deformation
of the corresponding Yukawa coupling, since the fits probe only the \CP-even directions. Hence the results
can be read as EFT constraints on the Wilson coefficients for these four light-quark operators, under the
assumption that other operators do not contribute.

\begin{table*}[!t]
    \centering
    \topcaption{
    Summary of the light-quark Yukawa coupling measurements for the $\PH\to\PZ\PZ\to4\Pell$ channel,
    showing central values with 68\% CL uncertainties and 95\% CL intervals (in square brackets).
    Results are obtained with all other light-quark couplings profiled simultaneously together with
    $\kappa_{\PZ}$ and $\kappa_{\PW}$.
    }
    \renewcommand{\arraystretch}{1.25}
   \begin{scotch}{llllll}
   $\kappa_\PQq$  &  Scenario & \multicolumn{2}{c}{Observed}   & \multicolumn{2}{c}{Expected} \\
   & &  \multicolumn{1}{c}{68\% \CL}  & \multicolumn{1}{c}{95\% \CL}  &  \multicolumn{1}{c}{68\% \CL}  & \multicolumn{1}{c}{95\% \CL}   \\
        \hline
        $\kappa_\PQu$ & All free& $(0.0 \pm 1.7) \times 10^3$ & $[-3.0, 3.0] \times 10^3$ & $(0.0^{+2.4}_{-2.6})\times 10^3$ & $[-4.3, 4.3] \times 10^3$ \\
        $\kappa_\PQu$ & $\abs{\kappa_\PV} \leq 1$ & $(0.0 \pm 1.4) \times 10^3$ & $[-2.1, 2.1] \times 10^3$ & $(0.0 \pm 1.7)\times 10^3$ & $[-2.5, 2.5] \times 10^3$ \\
        
        $\kappa_\PQd$ & All free & $(0.00 \pm 0.90)\times 10^3$ & $[-1.60, 1.60] \times 10^3$ & $(0.0 \pm 1.3) \times 10^3$ & $[-2.2, 2.2] \times 10^3$ \\
        $\kappa_\PQd$ & $\abs{\kappa_\PV} \leq 1$ & $(0.0 \pm 6.0)\times 10^2$ & $[-8.6, 8.6] \times 10^2$ & $(0.0 \pm 6.8) \times 10^2$ & $[-9.6, 9.5] \times 10^2$ \\
        
        $\kappa_\PQs$ & All free & $0^{+61}_{-64}$ & $[-98, 94] $ & $1^{+76}_{-81}$ & $[-116, 112] $ \\
        $\kappa_\PQs$ & $\abs{\kappa_\PV} \leq 1$ & $0^{+26}_{-28}$ & $[-38, 36] $ & $1^{+27}_{-30}$ & $[-40, 39] $ \\
        
        $\kappa_\PQc$ & All free & $-0.4 \pm 5.7$ & $[-8.6, 7.5]$ & $1.0^{+5.1}_{-7.9}$ & $[-9.6, 8.3]$ \\
        $\kappa_\PQc$ & $\abs{\kappa_\PV} \leq 1$ & $-0.3^{+2.3}_{-2.2}$ & $[-3.3, 2.8]$ & $1.0^{+1.2}_{-3.5}$ & $[-3.5, 2.9]$ \\
   \end{scotch}\label{table:lightYukawa}
\end{table*}

\begin{table*}[!htbp]
    \centering
    \topcaption{
    Summary of the light-quark Yukawa coupling measurements using the parameters
    $\barkappaq=\kappa_{\PQq} m_{\PQq}/m_{\PQb}$, where the values for $\kappa_{\PQq}$ 
    are taken from Table.~\ref{table:lightYukawa}, showing central values with 68\% CL
    uncertainties and 95\% CL intervals (in square brackets) for the $PH\to\PZ\PZ\to4\Pell$ channel.
    Results are obtained with all other light-quark couplings and
    $\kappa_{\PZ}$ and $\kappa_{\PW}$ profiled.
    }
    \renewcommand{\arraystretch}{1.25}
   \begin{scotch}{llllll}
        $\barkappaq$  &  Scenario  & \multicolumn{2}{c}{Observed}  & \multicolumn{2}{c}{Expected} \\
        & &  \multicolumn{1}{c}{68\% \CL}  & \multicolumn{1}{c}{95\% \CL}  &  \multicolumn{1}{c}{68\% \CL}  & \multicolumn{1}{c}{95\% \CL}   \\
        \hline
        $\barkappau$ & All free & $0.00 \pm 0.72$ & $[-1.32, 1.32]$ & $0.0^{+1.0}_{-1.1}$ & $[-1.85, 1.85]$ \\
        $\barkappau$ & $\abs{\kappa_\PV} \leq 1$ & $0.00 \pm 0.60$ & $[-0.91, 0.91]$ & $0.00 \pm 0.74$ & $[-1.06,1.06]$ \\
        $\barkappad$ & All free & $0.00^{+0.84}_{-0.85}$ & $[-1.51, 1.51]$ & $0.0\pm 1.2$ & $[-2.1, 2.1]$ \\
        $\barkappad$ & $\abs{\kappa_\PV} \leq 1$ & $0.00^{+0.56}_{-0.57}$ & $[-0.81, 0.81]$ & $0.00 \pm 0.64$ & $[-0.91, 0.90]$ \\
        $\barkappas$ & All free & $0.0\pm1.2$ & $[-1.9, 1.8]$ & $0.0^{+1.5}_{-1.6}$ & $[-2.3, 2.1]$ \\
        $\barkappas$ & $\abs{\kappa_\PV} \leq 1$ & $0.00^{+0.49}_{-0.53}$ & $[-0.73, 0.69]$ & $0.02^{+0.51}_{-0.58}$ & $[-0.77, 0.75]$ \\
        $\barkappac$ & All free & $-0.1 \pm 1.2$ & $[-1.9, 1.6]$ & $0.2^{+1.2}_{-1.7}$ & $[-2.1, 1.8]$ \\
        $\barkappac$ & $\abs{\kappa_\PV} \leq 1$ & $-0.06^{+0.49}_{-0.48}$ & $[-0.72, 0.61]$ & $0.22^{+0.22}_{-0.77}$ & $[-0.77, 0.64]$ \\
      \end{scotch}\label{table:lightYukawa_kappaBar}
\end{table*}

\section{Constraints on the Higgs boson self-coupling}\label{sec:kappa_lambda}

The Higgs boson self-interaction is parameterized by the modifier $\kappa_\lambda$, which rescales the term
in the Lagrangian of Eq.~(\ref{eq:kappa_lambda}). This modification is applied to both \ggH and
EW production modes. Corrections to the $\Pg\Pg$ LO process, taken from Ref.~\cite{Haisch:2021hvy}
and based on \mcfm, were first implemented in the \jhugen and \mela frameworks to enable event reweighting.
The same SMEFT formalism was then extended to EW production and implemented in \jhugen and \mela.

\begin{figure}[!bh]
    \centering
       \includegraphics[width=0.26\textwidth]{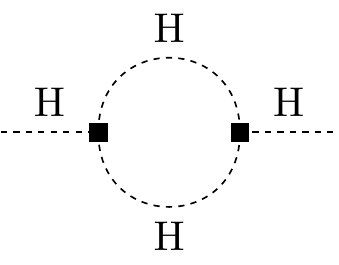}\\
       \includegraphics[width=0.36\textwidth]{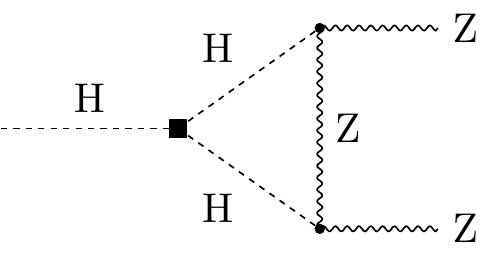}
     \caption{
     Feynman diagrams illustrating the loop-induced correction from the Higgs boson self-interaction in the propagator (upper)
     and in the $\PH\to\PZ\PZ$ decay (lower). Here, the black square represents the triple Higgs boson vertex.
      }\label{fig:diagrams_loops}
\end{figure}

\begin{figure*}[!th]
    \centering
    \includegraphics[width=0.45\textwidth]{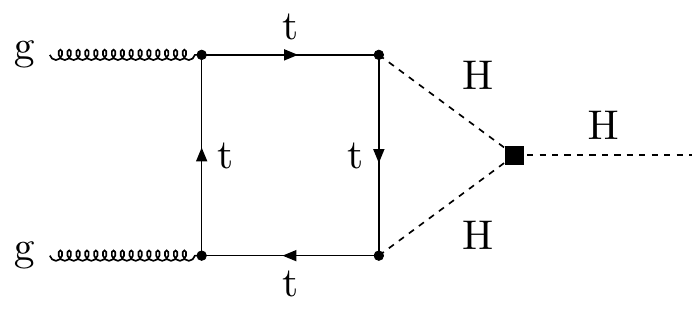}
    \includegraphics[width=0.42\textwidth]{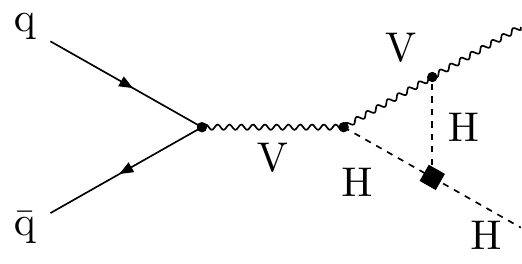} \\
    \includegraphics[width=0.45\textwidth]{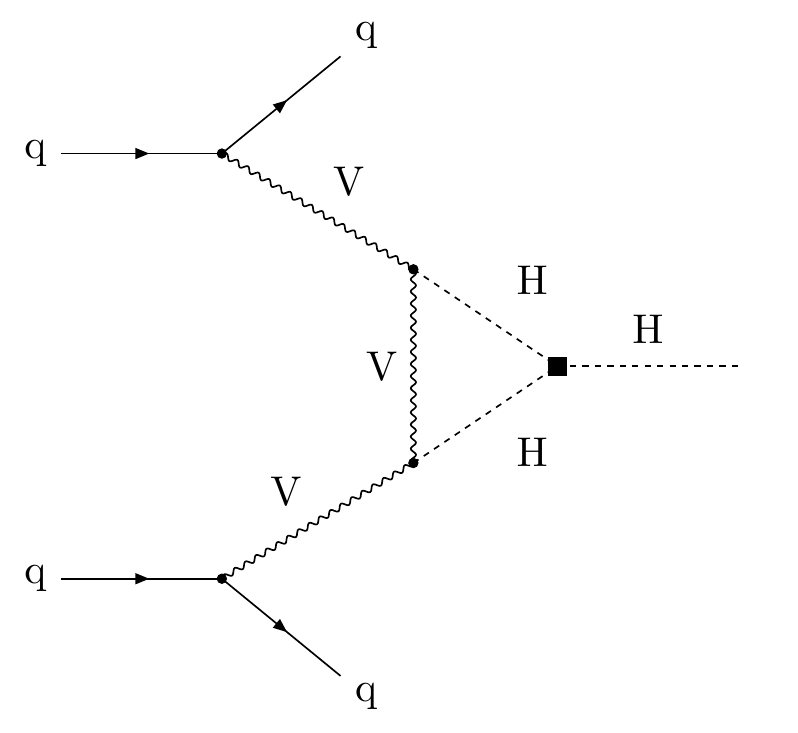}
    \includegraphics[width=0.45\textwidth]{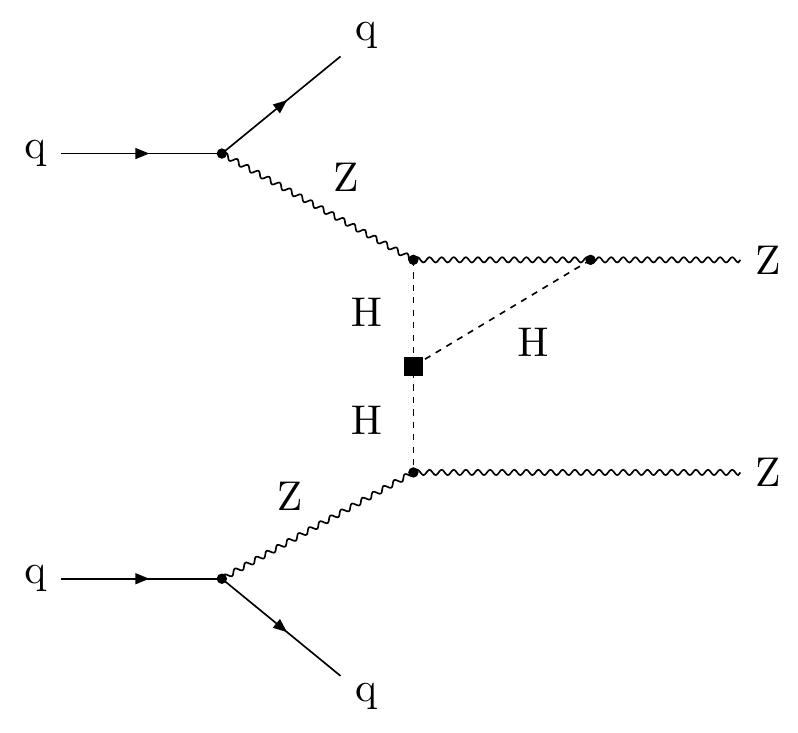}
    \caption{
    Feynman diagrams illustrating the correction from the Higgs boson self-interaction
    in the production processes considered in Fig.~\ref{fig:prod_modes}:
    \ggH (upper left), $\PV\PH$ associated production (upper right), $s$-channel VBF (lower left), and $t$-channel VBF (lower right). Here, the black
     square represents the triple Higgs boson vertex.
      }\label{fig:diagrams_c6}
\end{figure*}

Figures~\ref{fig:diagrams_loops} and~\ref{fig:diagrams_c6} show Feynman diagrams for the 
corrections modifying the Higgs boson propagator, 
its decay, and $\Pg\Pg$ and EW production processes.
The EW production corrections exhibit a topology similar to the decay correction, namely a triangle containing two Higgs bosons and one EW boson.
The difference is that in the decay the triangle is closed by a $\PZ$ boson, whereas in EW production it may
be closed by either a $\PZ$ or a $\PW$ boson (for example in $\PW\PW$ fusion). Therefore, the EW correction
has the same form as the decay correction in Ref.~\cite{Haisch:2021hvy}, with the EW leg being either $\PZ$ or $\PW$.

\begin{figure}[!htbp]
    \centering
    \includegraphics[width=\cmsFigWidthii]{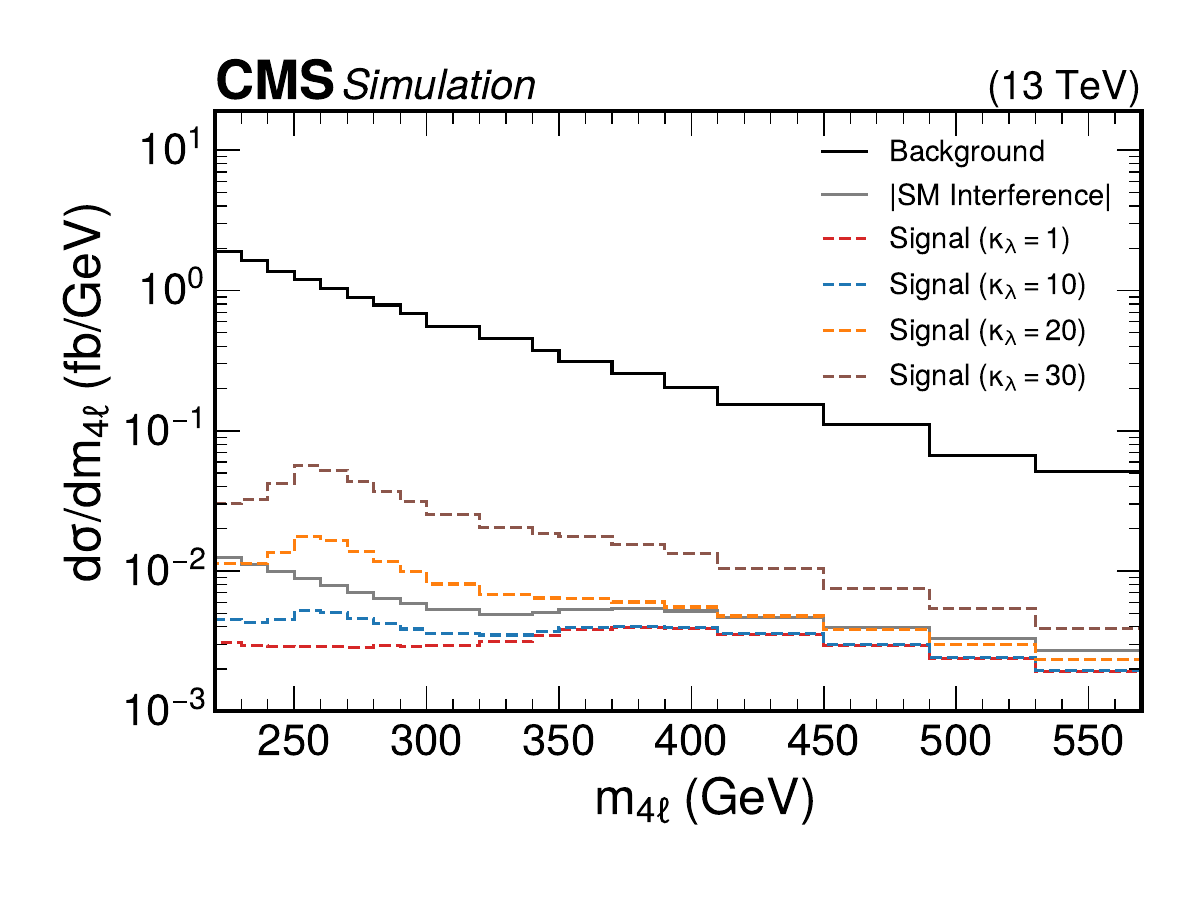}
    \caption{
    Distributions of \mell in simulation showing the background (black),
    the absolute value of the interference between the SM signal and background (gray),
    and the signal for various $\kappa_{\lambda}$ values: $\kappa_{\lambda}=1$ (red), 10 (blue), 20 (orange), and 30 (brown).
        }\label{fig:m4l_zoom_Inclusive}
\end{figure}

The $\kappa_\lambda$ modification leads to deviations in both the shape and the yield relative to the SM prediction,
producing an enhancement near $\mell=2m_\PH$, as shown in Fig.~\ref{fig:m4l_zoom_Inclusive}.
This feature originates from the two Higgs bosons in the loop becoming nearly on-shell,
and it is the dominant shape signature that distinguishes the modified Higgs boson self-coupling in
off-shell production from the SM contribution.
Introducing the SMEFT operator that changes $\kappa_\lambda$ also modifies \GH.
However, this effect is negligible in the off-shell region and is therefore omitted from the parameterization.

As in the other off-shell studies in this paper, the probability model is constructed from templates:
a Higgs boson signal term scaled by ${\mu_{\PH}}$,
its interference with the background scaled by $\sqrt{\mu_{\PH}}$, and the background, all constructed separately for the $\Pg\Pg$ and EW processes.

\begin{figure}[!th]
    \centering
    \includegraphics[width=\cmsFigWidth]{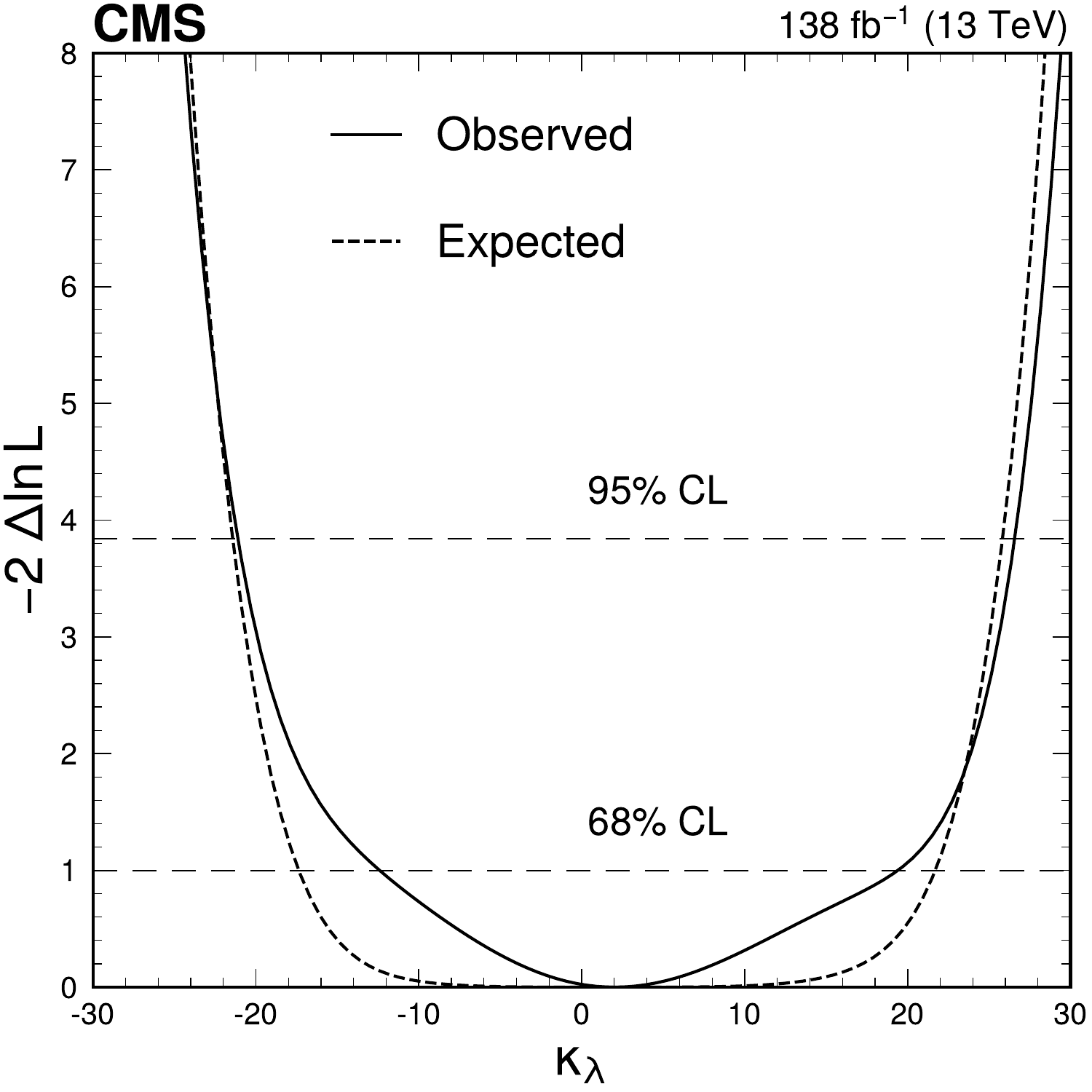}
    \caption{
    Observed (solid) and expected (dashed) profile likelihood scans from the fit for $\kappa_{\lambda}$.
    The black horizontal dashed lines indicate the 68 and 95\% CL thresholds.
    }\label{fig:scan_combined}
\end{figure}

The $\kappa_{\lambda}$ dependence of the signal and interference model can be separated into universal and non-universal contributions.
The universal contribution originates from wave-function renormalization (WFR) and enters through a second-order polynomial dependence on
$\kappa_{\lambda}$, while the non-universal terms arise from production and decay amplitude corrections, which are linear in $\kappa_{\lambda}$
at the amplitude level, and from propagator corrections, which introduce an additional second-order polynomial dependence.
Since the WFR contributions cancel at the amplitude level, the resulting $\kappa_{\lambda}$ dependence can be expressed as a quartic
polynomial in the rates.
A likelihood fit is performed with $\mu$ as a free parameter. The resulting scan is shown in Fig.~\ref{fig:scan_combined}, and
Table~\ref{tab:scan_klambda} lists the 68 and 95\% confidence intervals for both the expected and observed scans.
The observed limits are consistent with the expected sensitivity, and no significant deviation from the SM value of $\kappa_{\lambda}$ is found.
Although these constraints are weaker than those obtained from $\PH\PH$ or single $\PH$ on-shell production~\cite{2025139210},
they remain valuable because they are derived using different techniques and therefore offer complementary and independent information,
exploiting not only rate measurements but also shape information specific to the Higgs boson self-coupling.

In the SMEFT, $\kappa_{\lambda}$ is induced at LO by the dimension-6 operator
$\mathcal{O}_{6}=(H^{\dagger}H)^{3}$ with Wilson coefficient $c_{6}/\Lambda^2$, which alters the Higgs potential after EW
symmetry breaking and shifts the physical Higgs boson trilinear coupling. Neglecting other higher-dimension operators,
$\kappa_{\lambda}\simeq 1+\bar{c}_{6}$, where $\bar{c}_{6}=2c_6v^{4}/(m_\PH^2\Lambda^2)$.
Thus limits on $\kappa_{\lambda}$ can be translated into bounds on $\bar{c}_{6}$ under the single-operator assumption.

\begin{table}[!th]
    \centering
    \topcaption{
         Summary of the $\kappa_\lambda$ measurements, showing the 68\% CL
        (central values with uncertainties)
        and 95\%~\CL (in square brackets) intervals for the $\PH\to\PZ\PZ\to4\Pell$.
    }
    \renewcommand{\arraystretch}{1.25}
\begin{scotch}{llll}
   \multicolumn{2}{c}{Observed $\kappa_{\lambda}$} & \multicolumn{2}{c}{Expected $\kappa_{\lambda}$} \\
   \multicolumn{1}{c}{68\% \CL} & \multicolumn{1}{c}{95\% \CL} & \multicolumn{1}{c}{68\% \CL} & \multicolumn{1}{c}{95\% \CL} \\
\hline
  $2^{+17}_{-15}$ & $[-21,\,27]$ & $1^{+21}_{-18}$ & $[-21,\,26]$ \\
\end{scotch}\label{tab:scan_klambda}
\end{table}

\section{Measurement of the Higgs boson width, including BSM contributions}\label{sec:width}

The Higgs boson total width may differ from the SM expectation~\cite{deFlorian:2016spz} if it has decay modes
to as-yet-unknown BSM particles. The off-shell measurement, therefore, provides a probe for such invisible
or undetected decays utilizing the relationship in Eq.~(\ref{eq:offshell_xsec}) described in Section~\ref{sec:Introduction}. 
For instance,  Higgs boson decays to hypothetical dark-matter states would manifest
as an increased value of $\Gamma_{\PH}$. If these BSM states do not participate in the $\PH\to\PZ\PZ$
production or decay process, they will not bias the off-shell extraction method.
Instead their effect would appear simply as an increased value of $\Gamma_{\PH}$.

If BSM states contribute to $\PH\to\PZ\PZ$ production or decay, they can bias a simultaneous on- and
off-shell extraction of $\Gamma_{\PH}$ by modifying the differential distributions relative to the SM.
However, by including such effects as additional free parameters in the fit, they can be discovered
or constrained directly from the data.
Examples include new BSM particles in the \ggH loop or modified interactions of known states.
We show that this procedure causes only a modest loss of precision on $\Gamma_{\PH}$ unless a BSM
signal is actually observed. In other words, the off-shell method is ``no-lose'': either new BSM physics is revealed
(and the \GH measurement is affected because of that discovery), or the fit constrains \GH robustly
without bias or significant degradation in sensitivity.

One example of such a BSM effect affecting the $\PH\to\PZ\PZ$ production process, a hypothetical heavy
quark $\PQQ$ contributing to the \ggH loop, was tested in Ref.~\cite{CMS:2024eka} and discussed
in detail in Section~\ref{sec:H_heavy}. Another example of a BSM effect is the modification of the
tensor structure of the $\PH\PV\PV$ interactions, tested in Refs.~\cite{Sirunyan:2019twz,CMS:2022ley}.
Building on the approach of Refs.~\cite{Sirunyan:2019twz,CMS:2022ley,CMS:2024eka},
we take five progressive steps in the measurement of \GH:

\begin{enumerate}
    \item Baseline:
    While this study was previously performed on the same CMS data set in Ref.~\cite{CMS:2024eka},
    we optimized the analysis approach after recalibration of the CMS data and introducing improved
    observables via the \textsc{MiLoMerge} framework. This constitutes the baseline for the subsequent studies.
    The coupling modifiers $\kappa_{\PZ}=\kappa_{\PW}$ and $\kappa_{\PQt}$ are treated as free
    parameters to describe the Higgs boson interactions with the EW bosons and the quarks, respectively,
    with custodial symmetry imposed. Both $\kappa_{\PV}$ and $\kappa_{\PQt}$ are treated as positive parameters
    to mimic the effects of the signal strengths $\mu_\mathrm{F}$ and $\mu_\mathrm{V}$ from Ref.~\cite{CMS:2024eka}.

    \item No Custodial:
    Custodial symmetry was relaxed by allowing $\kappa_{\PZ}$ and $\kappa_{\PW}$ to vary freely and independently, in addition
    to the parameters described above. All free parameters are also allowed to take any real value,
    allowing the sign of interference with background to flip.

    \item Heavy BSM:
    The couplings $\widetilde{\kappa}_{\PQt}$, $\kappa_{\PQQ}$, and $\widetilde{\kappa}_{\PQQ}$
    discussed in Section~\ref{sec:H_heavy} were freely varied on top of the parameters already 
    varied when relaxing custodial symmetry.
    When considered alongside an unconstrained $\kappa_{\PQt}$, this provides a test for any heavy BSM particle
    entering the loop, allowing for either \CP-even or -odd interactions.

    \item Light Yukawa:
    The light-quark couplings $\kappa_{\PQu}$, $\kappa_{\PQd}$, $\kappa_{\PQs}$, and $\kappa_{\PQc}$,
    discussed in Section~\ref{sec:H_Yukawa}, are allowed to vary, 
    in addition to the parameters varied when custodial symmetry is relaxed.

    \item All: All of the parameters described above are allowed to vary simultaneously.
\end{enumerate}

\begin{figure}[!bh]
    \centering
    \includegraphics[width=\cmsFigWidth]{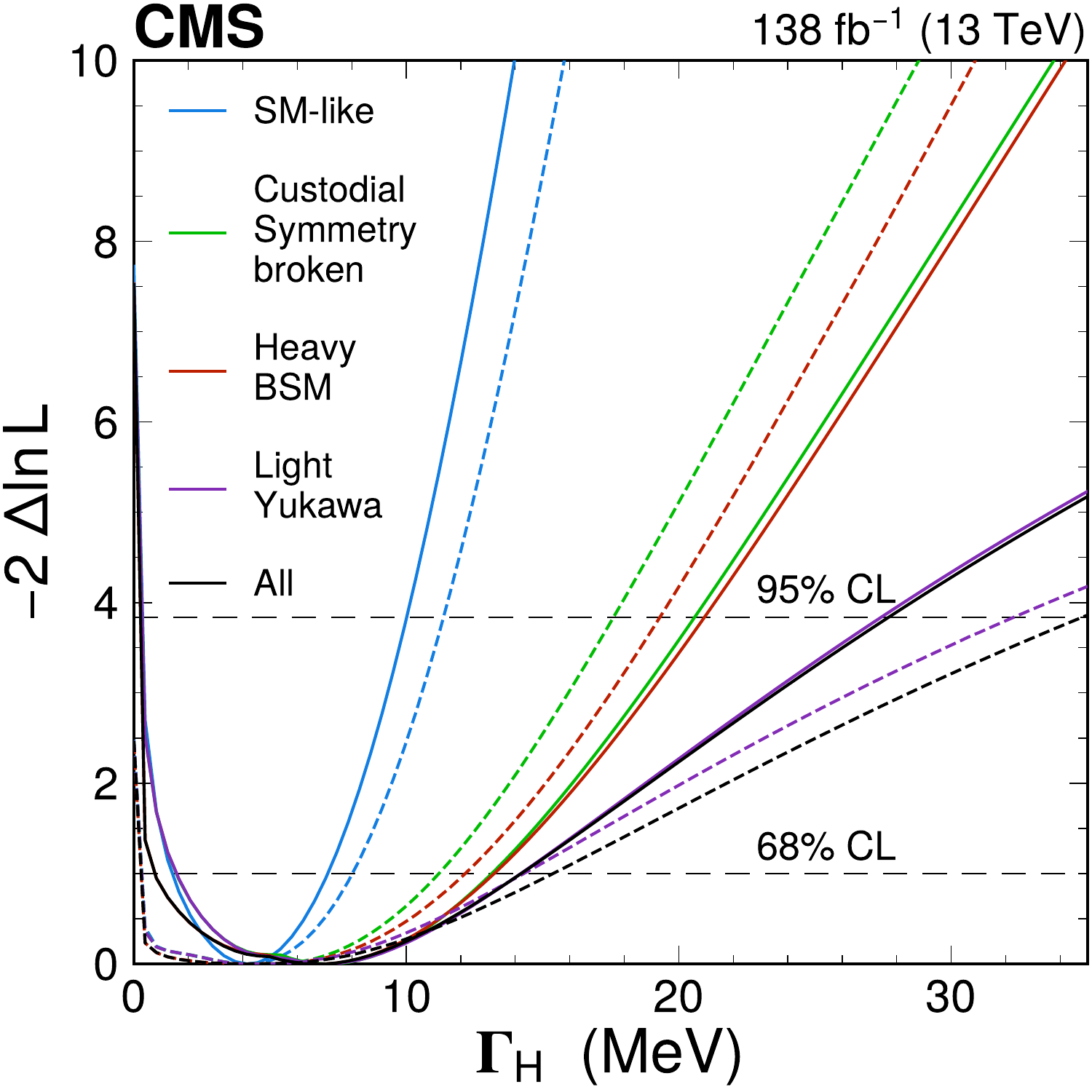}
    \caption{
Observed (solid) and expected (dashed) scans of $\Gamma_{\PH}$ for $\PH\to\PZ\PZ\to4\Pell$
under the five progressively inclusive BSM scenarios described in the text.
The black horizontal dashed lines indicate the 68 and 95\% CL thresholds.
    }\label{fig:width_scan}
\end{figure}

The parameterization implementing the steps described above is given in Eq.~(\ref{eq:off-shellYukawa})
and is discussed in detail in Sections~\ref{sec:H_heavy} and~\ref{sec:H_Yukawa}.
The on-shell parameterization follows Ref.~\cite{CMS:2025xkn}, since the effects of the parameters
listed above were available there. The value of \mH is fixed to $125.38\GeV$~\cite{CMS:2025xkn}.
Observed and expected constraints on \GH for each scenario are reported in Table~\ref{tab:width_table},
and the likelihood scans for $\Gamma_{\PH}$ are shown in Fig.~\ref{fig:width_scan}.

The baseline results are consistent with Ref.~\cite{CMS:2024eka}.
After breaking custodial symmetry, the limits widen, as expected, but the best fit value remains consistent. 
Allowing light-flavor $\kappa_\PQq$ parameters to vary weakens the constraints at larger $\GH$,
while varying heavy state parameters $\kappa_{\PQQ^\prime}$ loosens them at smaller $\GH$.
The final result has 95\% CL intervals that overlap those two fits. This demonstrates the robustness
of the off-shell method for determining $\GH$ when BSM contributions are profiled.

\begin{table*}[!t]
    \centering
    \topcaption{
    Summary of the measurements of the total \GH, showing the 68 and 95\% CL
    intervals for $\PH\to\PZ\PZ\to4\Pell$ under the five progressively more inclusive BSM scenarios discussed in text.
    }
    \renewcommand{\arraystretch}{1.25}
\begin{scotch}{llllll}
Scenario & \multicolumn{2}{c}{Observed $\Gamma_{\PH}$ (\MeV)} & & \multicolumn{2}{c}{Expected $\Gamma_{\PH}$ (\MeV)} \\
 & \multicolumn{1}{c}{68\% CL} & \multicolumn{1}{c}{95\% CL} & & \multicolumn{1}{c}{68\% CL} & \multicolumn{1}{c}{95\% CL} \\
\hline
Baseline & $4.2^{+2.9}_{-2.8}$ & $[0.3,\,10.0]$ & & $4.1 \pm 3.9$ & ${<}11.4$ \\
No Custodial & $7.0^{+6.1}_{-5.5}$ & $[0.3,\,20.6]$ & & $4.1^{+7.1}_{-3.9}$ & ${<}17.6$ \\
Heavy BSM & $6.6^{+6.6}_{-5.9}$ & $[0.3,\,21.0]$ & & $4.1^{+8.0}_{-3.9}$ & ${<}19.1$ \\
Light Yukawa & $7.0^{+7.2}_{-5.5}$ & $[0.3,\,27.4]$ & & $4.1^{+10.2}_{-3.8}$ & ${<}32.3$ \\
All & $7.0^{+7.2}_{-6.3}$ & $[0.3,\,27.7]$ & & $4.1^{+11.2}_{-3.9}$ & ${<}34.8$ \\
\end{scotch}\label{tab:width_table}
\end{table*}

\section{Measurement of the Higgs boson width from the combination of three channels}\label{sec:comb}

A combined measurement of \GH is obtained by fitting the on- and off-shell spectra
simultaneously across multiple channels, here including
on- and off-shell $\PH\to\PZ\PZ\to4\Pell$~\cite{CMS:2021nnc},
off-shell $\PH\to\PZ\PZ\to2\Pell2\Pgn$~\cite{CMS:2022ley},
and on- and off-shell $\PH\to\PW\PW\to2\Pell2\Pgn$~\cite{CMS:2026igg}.
Combining these channels increases sensitivity by adding complementary
event topologies, kinematic reach, and different background compositions: the $4\Pell$ channel provides excellent
mass resolution and powerful discriminants, while the $2\Pell2\Pgn$ channels contribute significantly at high
invariant mass due to the larger branching fraction.

The previous combination of the $\PH\to\PW\PW$ and $\PH\to\PZ\PZ$ channels occurred during 
Run-1~\cite{Khachatryan:2016ctc}. The two $\PH\to\PZ\PZ$ channels were combined using Run~2 data
in Ref.~\cite{CMS:2024eka}, but the Run~2 data set has since been reprocessed and the
$\PH\to\PZ\PZ\to4\Pell$ observables updated, as presented here. Therefore, we perform a new
combination that exploits all available $13\TeV$ data on \GH using the off-shell method,
assuming a SM-like evolution of the Higgs boson.

The fit uses channel-specific templates for signal, background, and interference, includes correlated systematic
uncertainties, and lets common physics parameters, $\Gamma_{\PH}$ and two signal strength modifiers,
$\mu_{\PV}$ and $\mu_{\mathrm{F}}$, are profiled. Here, $\mu_{\PV}$ denotes the signal strength for EW-induced
production, (e.g. VBF and $\PV\PH$) and $\mu_{\mathrm{F}}$ denotes the signal strength for QCD-induced production
(\eg \ggH). This combined treatment tightens the constraint on $\Gamma_{\PH}$ compared with any
single channel.

\begin{figure}[!th]
    \centering
    \includegraphics[width=0.45\textwidth]{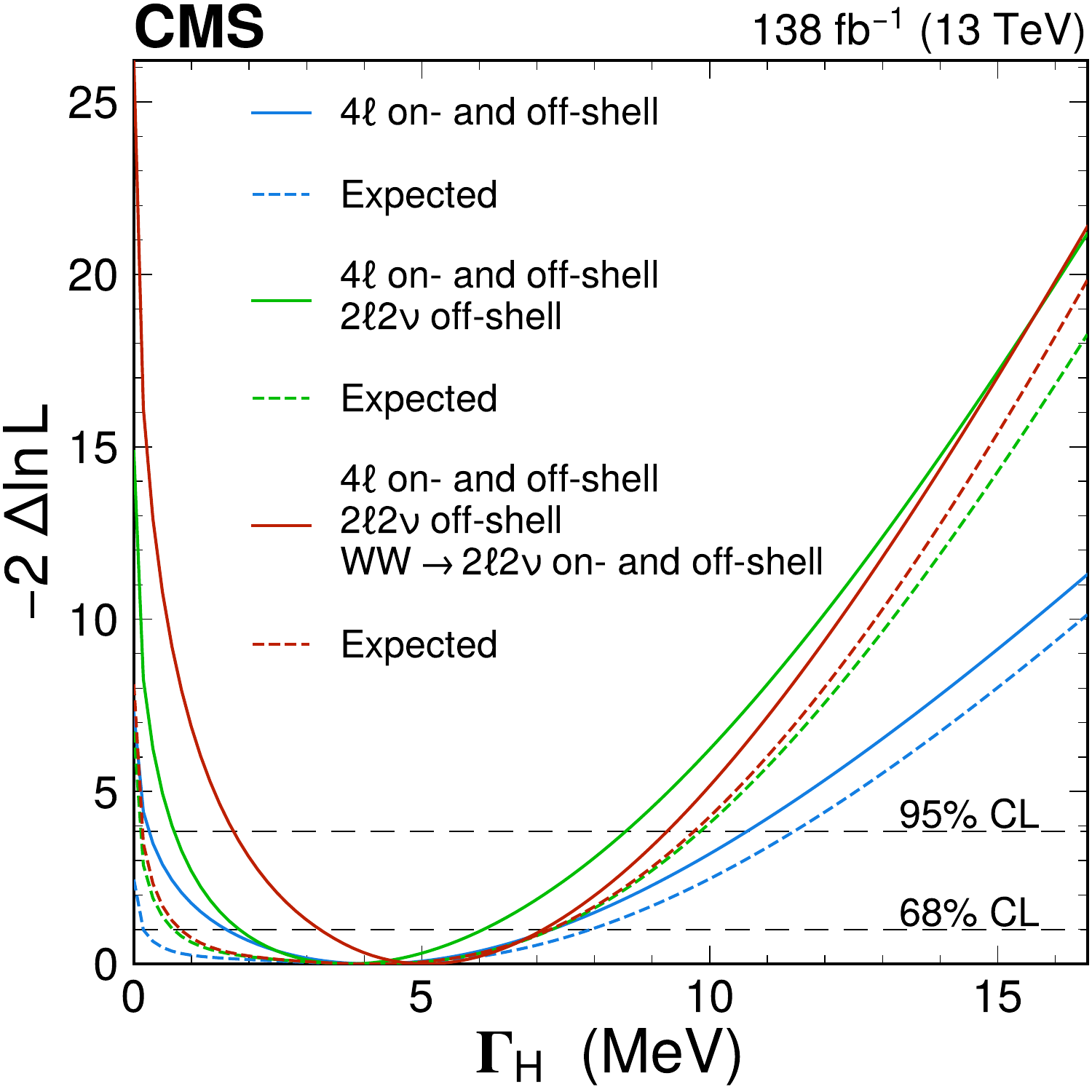}
    \includegraphics[width=0.45\textwidth]{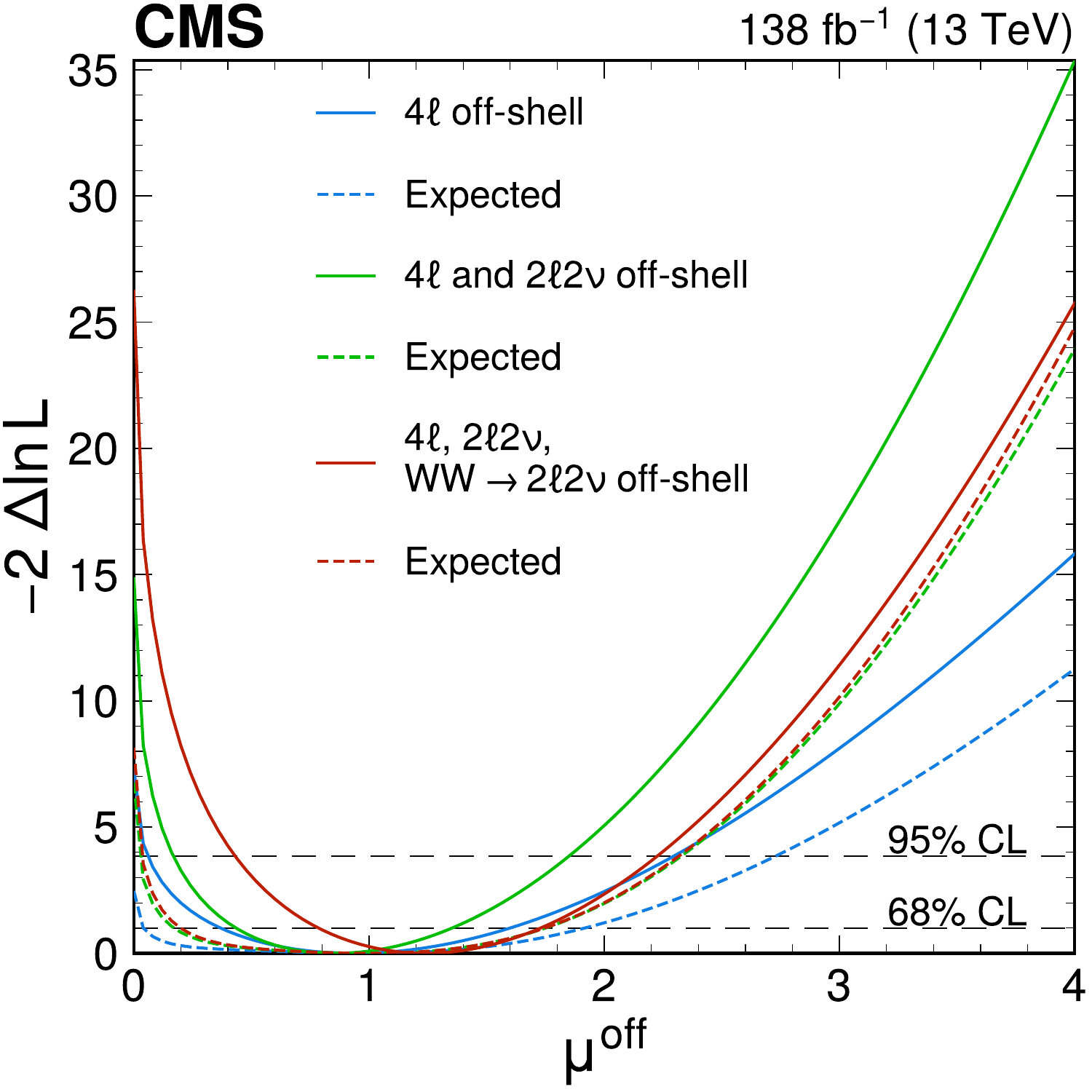}
    \caption{
    \cmsLLeft: Observed (solid) and expected (dashed) scans of \GH for
    on- and off-shell $\PH\to\PZ\PZ\to4\Pell$,
    and in combination with off-shell $\PH\to\PZ\PZ\to2\Pell2\Pgn$,
    and with on- and off-shell $\PH\to\PW\PW\to2\Pell2\Pgn$.
    \cmsRRight: The off-shell signal strength, $\mu^{\text{off}}$.
    The black horizontal dashed lines mark the 68 and 95\% CL thresholds.
    }\label{fig:combination_gammaH}
\end{figure}

The combination results are shown in Fig.~\ref{fig:combination_gammaH} (left) and Table~\ref{table:combination_gammaH}.
Studies using simulated pseudo-experiments indicate that the deviations between the expected and observed results are consistent
with statistical fluctuations in both the individual measurements~\cite{CMS:2026igg,CMS:2024eka,CMS:2022ley}
and the combination. Such an effect occurs when the deficits in signal-enriched
regions favor scenarios with greater destructive interference.

An alternative parameterization of the signal process is to introduce separate signal strengths in the on- and off-shell regions: 
$\mu^{\text{on}}$ and $\mu^{\text{off}}$, respectively, instead of using \GH and a single signal strength.
These quantities can be related as $\mu^{\text{on}} = \mu^{\text{off}} \,\GH^\text{SM}/\GH$, following Eq.~(\ref{eq:offshell_xsec}).
The $\mu^{\text{off}}$ parameter was introduced in Eq.~(\ref{eq:poffshell}), where we assign 
the same value for both EW-induced and QCD-induced production. 
The two parameterizations are equivalent at the points $\Gamma_\PH = 0$ and $\mu^\text{off} = 0$.
The combined measurements of $\mu^{\text{off}}$ are shown in Fig.~\ref{fig:combination_gammaH} (right) and 
Table~\ref{table:combination_muoffshell}. 
The significance for excluding the null hypothesis of no off-shell Higgs boson production, 
corresponding to $\mu^{\text{off}}=0$, 
exceeds 5 standard deviations, compared with an expected significance of 2.9 standard deviations,
as seen in Fig.~\ref{fig:combination_gammaH}.

\begin{table*}[!th]
    \centering
    \topcaption{
    Summary of the measurements of the total \GH, showing the 68 and 95\% CL
    intervals for on- and off-shell $\PH\to\PZ\PZ\to4\Pell$,
    combination with off-shell $\PH\to\PZ\PZ\to2\Pell2\Pgn$,
    and with on- and off-shell $\PH\to\PW\PW\to2\Pell2\Pgn$.
    }
    \renewcommand{\arraystretch}{1.25}
\begin{scotch}{llllll}
Scenario & \multicolumn{2}{c}{Observed $\Gamma_{\PH}$ (\MeV)} & & \multicolumn{2}{c}{Expected $\Gamma_{\PH}$ (\MeV)} \\
 & \multicolumn{1}{c}{68\% CL} & \multicolumn{1}{c}{95\% CL} & & \multicolumn{1}{c}{68\% CL} & \multicolumn{1}{c}{95\% CL} \\
\hline
$\PH\to\PZ\PZ\to4\Pell$ & $4.1^{+3.1}_{-2.5}$ & $[0.3,\,10.6]$ & & $4.1^{+3.8}_{-4.0}$ & ${<}11.5$ \\
with  $\PH\to\PZ\PZ\to2\Pell2\Pgn$ & $3.7^{+2.4}_{-1.9}$ & $[0.7,\,8.5]$ & & $4.1^{+3.1}_{-3.5}$ & $[0.1,\,9.8]$ \\
with  $\PH\to\PZ\PZ / \PW\PW\to2\Pell2\Pgn$ & $5.1^{+2.0}_{-1.8}$ & $[1.7,\,9.3]$ & & $4.1^{+3.1}_{-3.3}$ & $[0.2,\,9.7]$ \\
\end{scotch}\label{table:combination_gammaH}
\end{table*}

\begin{table}[!th]
    \centering
    \topcaption{
    Summary of the combined measurement for $\mu^\text{off}$, showing the 68 and 95\% CL
    intervals for on- and off-shell $\PH\to\PZ\PZ\to4\Pell$ in
    combination with off-shell $\PH\to\PZ\PZ\to2\Pell2\Pgn$ and on- and off-shell $\PH\to\PW\PW\to2\Pell2\Pgn$.
    }
    \renewcommand{\arraystretch}{1.25}
    \begin{scotch}{llll}
    \multicolumn{2}{c}{Observed $\mu^\text{off}$} & \multicolumn{2}{c}{Expected $\mu^\text{off}$} \\
    \multicolumn{1}{c}{68\% CL} & \multicolumn{1}{c}{95\% CL} & \multicolumn{1}{c}{68\% CL} & \multicolumn{1}{c}{95\% CL} \\
    \hline
    $1.24^{+0.49}_{-0.45}$ & $[0.43,\, 2.22]$ & $1.00^{+0.75}_{-0.80}$ & $[0.04,\, 2.32]$ \\
    \end{scotch}\label{table:combination_muoffshell}
\end{table}

\section{Summary}\label{sec:summary}
Using a data set of proton-proton collisions at $\sqrt{s}=13\TeV$, corresponding to an integrated luminosity of 138\fbinv,
and improved observables, a unified study of Higgs boson production that exploits both on-shell and off-shell regions
in $\PH\to\PZ\PZ$ decays was performed. The off-shell region provides distinctive shape information that complements rate
measurements, and combining on- and off-shell data allows for a range of physics scenarios with reduced
model dependence to be probed.

Four classes of effects were tested.
First, Higgs boson compositeness was probed through a $q^{2}$-dependent form factor
(where $q^{2}$ is the squared four-momentum transfer of the Higgs boson),
and a lower limit of $\Lambda_{\PH} > 870\GeV$ was set on the compositeness scale at 95\% confidence level.
Second, possible heavy colored states in the gluon-fusion loop were tested by fitting \CP-even and \CP-odd
modifications of the top and a generic heavy quark ($\kappa_{\PQt},\kappa_{\PQQ},\widetilde{\kappa}_{\PQt},\widetilde{\kappa}_{\PQQ}$).
These limits can be presented as direct bounds on the couplings to the Higgs boson or translated into constraints 
on the corresponding standard model effective field theory (SMEFT) Wilson coefficients. 
Third, by combining on- and off-shell measurements, \CP-even deviations of the light-quark Yukawa
couplings ($\kappa_{\PQu},\kappa_{\PQd},\kappa_{\PQs},\kappa_{\PQc}$) were constrained,
where the off-shell information is crucial because 
it allows for relaxed on-shell assumptions (for example $\abs{\kappa_\PV}\leq 1$) while retaining meaningful bounds.
Fourth, off-shell kinematic information was utilized to probe
the Higgs boson trilinear coupling $\kappa_{\lambda}$ via the SMEFT operator $\mathcal{O}_{6}=(H^{\dagger}H)^{3}$.
The limits are complementary to on-shell and double-H searches because they exploit different
kinematic shape signatures.

Finally, the total Higgs boson width \GH was investigated under progressively more general BSM hypotheses:
relaxing custodial symmetry, profiling the heavy- and light-sector couplings separately, and then profiling all parameters. 
The off-shell method was found to be robust: introducing additional BSM degrees of freedom either reveals
a direct BSM signature or leaves a strong constraint on $\Gamma_{\PH}$ with only modest loss of precision.
A new $\sqrt{s}=13\TeV$ combination including the $\PH\to\PZ\PZ$ and $\PH\to\PW\PW$ channels was performed to maximize
sensitivity to $\Gamma_{\PH}$, yielding $\Gamma_{\PH}=5.1^{+2.0}_{-1.8}\MeV$ and a 95\% confidence-level 
interval of $1.7$ to $9.4\MeV$.
The hypothesis of no off-shell Higgs boson production was excluded at the $5$ standard deviation level.

Taken together, these measurements showed that off-shell analyses substantially broaden the Higgs boson program:
they deliver shape-based sensitivity to substructure, loop effects, light-quark Yukawa couplings, and the self-coupling, and
they produce a BSM-aware and precise combined constraint on the total width of the Higgs boson.

\begin{acknowledgments}

We congratulate our colleagues in the CERN accelerator departments for the excellent performance of the LHC and thank the technical and administrative staffs at CERN and at other CMS institutes for their contributions to the success of the CMS effort. In addition, we gratefully acknowledge the computing centers and personnel of the Worldwide LHC Computing Grid and other centers for delivering so effectively the computing infrastructure essential to our analyses. Finally, we acknowledge the enduring support for the construction and operation of the LHC, the CMS detector, and the supporting computing infrastructure provided by the following funding agencies: SC (Armenia), BMFWF and FWF (Austria); FNRS and FWO (Belgium); CNPq, CAPES, FAPERJ, FAPERGS, and FAPESP (Brazil); MES and BNSF (Bulgaria); CERN; CAS, MoST, and NSFC (China); MINCIENCIAS (Colombia); MSES and CSF (Croatia); RIF (Cyprus); SENESCYT (Ecuador); ERC PRG and PSG, TARISTU24-TK10 and MoER TK202 (Estonia); Academy of Finland, MEC, and HIP (Finland); CEA and CNRS/IN2P3 (France); SRNSF (Georgia); BMFTR, DFG, and HGF (Germany); GSRI (Greece); MATE and NKFIH (Hungary); DAE and DST (India); IPM (Iran); SFI (Ireland); INFN (Italy); MSIT and NRF (Republic of Korea); MES (Latvia); LMTLT (Lithuania); MOE and UM (Malaysia); BUAP, CINVESTAV, CONACYT, LNS, SEP, and UASLP-FAI (Mexico); MOS (Montenegro); MBIE (New Zealand); PAEC (Pakistan); MSHE, NSC, and NAWA (Poland); FCT (Portugal);  MESTD (Serbia); MICIU/AEI and PCTI (Spain); MOSTR (Sri Lanka); Swiss Funding Agencies (Switzerland); MST (Taipei); MHESI (Thailand); TUBITAK and TENMAK (T\"{u}rkiye); NASU (Ukraine); STFC (United Kingdom); DOE and NSF (USA).

\begin{sloppypar}
\setlength\emergencystretch{\hsize}
\hyphenation{Rachada-pisek} Individuals have received support from the Marie-Curie program and the European Research Council and Horizon 2020 Grant, contract Nos.\ 675440, 724704, 752730, 758316, 765710, 824093, 101115353, 101002207, 101001205, and COST Action CA16108 (European Union); the Leventis Foundation; the Alfred P.\ Sloan Foundation; the Alexander von Humboldt Foundation; the Science Committee, project no. 22rl-037 (Armenia); the Fonds pour la Formation \`a la Recherche dans l'Industrie et dans l'Agriculture (FRIA) and Fonds voor Wetenschappelijk Onderzoek contract No. 1228724N (Belgium); the Beijing Municipal Science \& Technology Commission, No. Z191100007219010, the Fundamental Research Funds for the Central Universities, the Ministry of Science and Technology of China under Grant No. 2023YFA1605804, the Natural Science Foundation of China under Grant No. 12535004, and USTC Research Funds of the Double First-Class Initiative No.\ YD2030002017 (China); the Ministry of Education, Youth and Sports (MEYS) of the Czech Republic; the Shota Rustaveli National Science Foundation (Georgia); the Deutsche Forschungsgemeinschaft (DFG), among others, under Germany's Excellence Strategy -- EXC 2121 ``Quantum Universe" -- 390833306, and under project number 400140256 - GRK2497; the Hellenic Foundation for Research and Innovation (HFRI), Project Number 2288 (Greece); the Hungarian Academy of Sciences, the New National Excellence Program - \'UNKP, the NKFIH research grants K 131991, K 138136, K 143460, K 143477, K 147557, K 146913, K 146914, K 147048, TKP2021-NKTA-64, and 2025-1.1.5-NEMZ\_KI-2025-00004, and MATE KKP and KKPCs Research Excellence and Flagship Research Groups grants (Hungary); the Council of Science and Industrial Research, India; ICSC -- National Research Center for High Performance Computing, Big Data and Quantum Computing, FAIR -- Future Artificial Intelligence Research, and CUP I53D23001070006 (Mission 4 Component 1), funded by the NextGenerationEU program, the Italian Ministry of University and Research (MUR) under Bando PRIN 2022 -- CUP I53C24002390006, PRIN PRIMULA 2022RBYK7T (Italy); the Latvian Council of Science; the Ministry of Science and Higher Education, project no. 2022/WK/14, and the National Science Center, contracts Opus 2021/41/B/ST2/01369, 2021/43/B/ST2/01552, 2023/49/B/ST2/03273, and the NAWA contract BPN/PPO/2021/1/00011 (Poland); the Funda\c{c}\~ao para a Ci\^encia e a Tecnologia (Portugal); the National Priorities Research Program by Qatar National Research Fund; MICIU/AEI/10.13039/501100011033, ERDF/EU, ``European Union NextGenerationEU/PRTR", projects PID2022-142604OB-C21, PID2022-139519OB-C21, PID2023-147706NB-I00, PID2023-148896NB-I00, PID2023-146983NB-I00, PID2023-147115NB-I00, PID2023-148418NB-C41, PID2023-148418NB-C42, PID2023-148418NB-C43, PID2023-148418NB-C44, PID2024-158190NB-C22, RYC2021-033305-I, RYC2024-048719-I, CNS2023-144781, CNS2024-154769 and Plan de Ciencia, Tecnolog{\'i}a e Innovaci{\'o}n de Asturias, Spain; the Chulalongkorn Academic into Its 2nd Century Project Advancement Project, the National Science, Research and Innovation Fund program IND\_FF\_68\_369\_2300\_097, and the Program Management Unit for Human Resources \& Institutional Development, Research and Innovation, grant B39G680009 (Thailand); the Eric \& Wendy Schmidt Fund for Strategic Innovation through the CERN Next Generation Triggers project under grant agreement number SIF-2023-004; the Kavli Foundation; the Nvidia Corporation; the SuperMicro Corporation; the Welch Foundation, contract C-1845; Advanced Research Computing at Hopkins (ARCH) under NSF grant OAC1920103 (USA); and the Weston Havens Foundation (USA).
\end{sloppypar}
\end{acknowledgments}\section*{Data availability} Release and preservation of data used by the CMS Collaboration as the basis for publications is guided by the  \href{https://doi.org/10.7483/OPENDATA.CMS.1BNU.8V1W}{CMS data preservation, re-use and open access policy}.

\bibliography{auto_generated} 

\cleardoublepage \appendix\section{The CMS Collaboration \label{app:collab}}\begin{sloppypar}\hyphenpenalty=5000\widowpenalty=500\clubpenalty=5000\cmsinstitute{Yerevan Physics Institute, Yerevan, Armenia}
{\tolerance=6000
A.~Gevorgyan\cmsorcid{0000-0003-2751-9489}, A.~Hayrapetyan, V.~Makarenko\cmsorcid{0000-0002-8406-8605}, A.~Tumasyan\cmsAuthorMark{1}\cmsorcid{0009-0000-0684-6742}
\par}
\cmsinstitute{Institut f\"{u}r Hochenergiephysik, Vienna, Austria}
{\tolerance=6000
P.S.~Hussain\cmsorcid{0000-0002-4825-5278}, M.~Sonawane\cmsorcid{0000-0003-0510-7010}
\par}
\cmsinstitute{Marietta Blau Institute for Particle Physics, Vienna, Austria}
{\tolerance=6000
W.~Adam\cmsorcid{0000-0001-9099-4341}, L.~Benato\cmsorcid{0000-0001-5135-7489}, T.~Bergauer\cmsorcid{0000-0002-5786-0293}, M.~Dragicevic\cmsorcid{0000-0003-1967-6783}, S.~Gundacker\cmsorcid{0000-0003-2087-3266}, A.K.~Guven\cmsorcid{0009-0004-5670-5138}, M.~Jeitler\cmsAuthorMark{2}\cmsorcid{0000-0002-5141-9560}, N.~Krammer\cmsorcid{0000-0002-0548-0985}, A.~Li\cmsorcid{0000-0002-4547-116X}, D.~Liko\cmsorcid{0000-0002-3380-473X}, M.~Matthewman, A.~Pfeiffer\cmsorcid{0000-0001-5328-448X}, J.~Schieck\cmsAuthorMark{2}\cmsorcid{0000-0002-1058-8093}, R.~Sch\"{o}fbeck\cmsAuthorMark{2}\cmsorcid{0000-0002-2332-8784}, M.~Shooshtari\cmsorcid{0009-0004-8882-4887}, N.~Van~Den~Bossche\cmsorcid{0000-0003-2973-4991}, W.~Waltenberger\cmsorcid{0000-0002-6215-7228}, C.E.~Wulz\cmsAuthorMark{2}\cmsorcid{0000-0001-9226-5812}
% ...
\par}
$^{1}$Also at Yerevan State University, Yerevan, Armenia\\
$^{2}$Also at Technische Universit\"{a}t Wien, Vienna, Austria\\
% ...
\end{sloppypar}
\end{document}